\documentclass[useAMS, usenatbib]{mn2e}
\usepackage[dvips]{graphicx}
\input{epsf}

\title[ImpZ: a new photometric redshift code]{ImpZ: a new photometric redshift code for galaxies and quasars}
\author[T.S.R Babbedge et al.]
       {T.S.R Babbedge,$^1$\thanks{Email : tsb1@imperial.ac.uk} M. Rowan--Robinson, $^1$ E. Gonzalez-Solares,$^2$ \newauthor M. Polletta,$^3$
       S. Berta,$^4$ I. P\'{e}rez--Fournon,$^5$ S. Oliver,$^6$ \newauthor D.M. Salaman,$^6$ M. Irwin,$^2$ S.J. Weatherley.$^1$\\
        $^1$Astrophysics Group, Blackett Laboratory, Imperial College London, Prince Consort Road, London SW7 2BW, UK.\\
        $^2$Institute of Astronomy, University of Cambridge, Madingley Road, Cambridge CB3 0HA, UK.\\
        $^3$Infrared Processing and Analysis Centre, California Institute of Technology, 100--22, Pasadena, CA 91125.\\
        $^4$Dipartimento di Astronomia, Vicolo dell'Osservatorio 5, I--35122 Padua, Italy.\\
        $^5$Instituto de Astrofisica de Canarias, 38200 La Laguna, Tenerife, Spain.\\
        $^6$Astronomy Centre, CPES, University of Sussex, Falmer, Brighton BN1 9QJ, UK.}
\date{Accepted 0000 January 00.
      Received 0000 January 00;
      in original form 0000 January 00}

\pagerange{\pageref{firstpage}--\pageref{lastpage}}
\pubyear{2004}
\begin{document}
\maketitle
\label{firstpage}

\begin{abstract}
We present a combined galaxy-quasar approach to template-fitting photometric redshift techniques and show the method to be a powerful one.\\
The code, ImpZ, is presented, developed and applied to two spectroscopic redshift catalogues, namely the Isaac Newton Telescope Wide Angle Survey ELAIS N1 and N2 fields and the Chandra Deep Field North. In particular, optical size information is used to improve the redshift determination.  The success of the code is shown to be very good with $\Delta z/(1+z)$ constrained to within 0.1 for 92 per cent of the galaxies in our sample.\\
The extension of template-fitting to quasars is found to be reasonable with $\Delta z/(1+z)$ constrained to within 0.25 for 68 per cent of the quasars in our sample. Various template extensions into the far-UV are also tested.
%The extension of template-fitting to quasars is found to be reasonable with 52 per cent of sources contstrained to 25 per cent of their spectroscopic redshift. Various template extensions into the far-UV are also tested.
\end{abstract}


\newcommand{\mnras}{MNRAS}
\newcommand{\apj}{ApJ}
\newcommand{\apjl}{ApJL}
\newcommand{\apjs}{ApJS}
\newcommand{\aj}{AJ}
\newcommand{\aap}{AAP}
\newcommand{\araa}{ARA\&A}
\newcommand{\pasp}{PASP}
\newcommand{\nat}{Nature}

%%%%%%%%%%%%%%%%%%%%%%%%%%%%%%%%%%%%%%%%%%%%%%%%%%%%%%%%%%%%%%%%%%%%%%%
\section{Introduction}
\label{sec:intro}
%SIRTF will be renamed in DECEMBER!!!!!
Photometric redshifts are a powerful statistical tool for studies of the evolutionary properties of galaxies, in particular of faint galaxies, for which spectroscopic data is hard or impossible to obtain.

Photometric redshifts are faster to measure than spectroscopic redshifts and can be applied to much fainter magnitudes since the bin sizes are larger ($\approx$1000{\rm \AA}  vs. 1--2{\rm \AA}).  There is however a trade-off with redshift precision -- \cite{Hogg1998AJ....115.1418H} found that photometric redshifts can be predicted with an accuracy of $\Delta z\approx$0.1 (0.3) in $z$ for 66 per cent (99 per cent) of the sources examined.

%Deriving redshifts from photometry dates back to \cite{Baum1962IAUS...15..390B} who used 9--band photoelectric data to locate the 4000{\rm \AA} Balmer break in elliptical galaxies.  \cite{Loh1986ApJ...303..154L} generalised this technique with 6--band photometry applied to a wide range of galaxy types.  \cite{Koo1985AJ.....90..418K} used 4 photographic bands to estimate redshift from redshift contours on colour--colour plots.\\
The bulk of the photometric redshift identification is carried out using the broadband continuum shape and presence and/or absence of spectral breaks (like the $4000{\rm \AA}$ Balmer Break) or the onset of the Ly$\alpha$ forest and Lyman limit which enter optical wavebands at high redshift (the Ly$\alpha$ forest effect, e.g. \citealt{Madau1995ApJ...441...18M} and \citealt{Steidel1987ApJ...313..171S}).  Although methods based on training sets such as polynomial fitting (e.g. \citealt*{Wang1998AJ....116.2081W}) and artificial neural networks (e.g \citealt{Ball2004MNRAS.348.1038B} and \citealt{Tagliaferri2002astroph}) have had some success in determining photometric redshifts, we decide to utilise the template-fitting procedure (but with an element of training in that the templates have been adapted to improve fits).   This is because there is a relative paucity of galaxies with both spectroscopic redshifts and photometry in the bandpasses used in the catalogues under study here, particularly at higher redshifts, making the construction of a training set unfeasible.  Empirical methods are also hard to apply outside of the boundaries in which they were defined -- such as the redshift distribution of the training set or the photometric bands use.

In the template-fitting method (e.g. \citealt*{Sawicki1997AJ....113....1S}; \citealt{Giallongo1998AJ....115.2169G}) the observed galaxy fluxes, $f_{i}^{obs}$ in the $i^{th}$ band, are compared to a library of reference fluxes, f$_{i}^{templ}$(z,T), where T is a set of parameters that account for the template galaxy's morphological type, age, metallicity and dust.  We then fit the observed fluxes to the library fluxes using $\chi^2$ minimisation.  As well as deriving redshifts, the procedure produces information on spectral (template) type, although this is less robust due to the degeneracies in the parameter space (see \S\ref{subsubsec:degeneracy}).  A way of breaking these degeneracies is to use Bayesian probability (e.g. \citealt{Jaynes2000}) to weight the solutions based on a prior knowledge of the expected population distributions.  Application of Bayesian methods to photometric redshifts has been presented in, for example, \cite{Kodama1999MNRAS.302..152K} and \cite{Benitez2000ApJ...536..571B}.  Usually such applications use priors such as the expected redshift distribution of the sample but this naturally suppresses unbiased information on the true redshift distribution.  Here Bayesian methods are implemented using absolute magnitude limits and extinction distributions which it is hoped retain the power of Bayesian methods without unduly influencing the underlying science (see \S\ref{subsec:code}).\\

One important refinement is the consideration of extinction for galaxies.  \cite{Madau1999} used a single dust absorption correction of A$_{1500}$=1.2 mag, except for galaxies in the redshift range 0.75--1.75 where the equivalent extinction at 2800{\rm \AA}  was used.  Galaxy evolution models such as those of \cite{LeBorgne2002A&A...386..446L} include evolution of dust extinction with time.  In order to allow for variation in extinction from galaxy to galaxy, extinction needs to be solved as an additional free parameter to redshift.  In \cite{Steidel1999ApJ...519....1S} dust absorption was corrected for by assuming that colour deviations in their sample galaxies were entirely due to dust absorption, based on the empirical relation between far-infrared emission and the observed UV spectrum slope, as derived by \cite{Meurer1999ApJ...521...64M}.  More recently, \cite*{Thompson2001ApJ...546..694T} used Spectral Energy Distribution (SED) template-fitting photometric redshift techniques in the deep NICMOS northern HDF, fitting extinction as a parameter ranging from $E(B-V)$=0 to 1.0.  The study of \cite*{Bolzonella2000A&A...363..476B} found that the inclusion of A$_{v}$ as a free parameter caused significant increases in aliasing.  In a similar technique developed in \cite{RR2003MNRAS.345..819R}, hereafter RR03, these aliasing problems were reduced by setting several A$_{v}$ priors (see \S\ref{subsec:code}).\\

In this paper the SED template-fitting set out in RR03 is refined and applied to two spectroscopic redshift galaxy samples from the European Large--Area $ISO$ Survey (ELAIS; \citealt{Oliver2000MNRAS.316..749O}) N1 and N2 fields of the Isaac Newton Telescope Wide--Angle Survey (INT WAS; \citealt{Mcmahon2001NewAR..45...97M}), a part of the INT Wide--Field Survey (INT WFS), and also the Chandra Deep Field North, CDFN) for validation purposes.  The effect of non-zero A$_{v}$, different SED templates and various template extensions into the far-UV are explored, as is the inclusion of several different quasar templates and the applicability of template-fitting techniques to quasar-like sources.  In \S\ref{sec:method} the photometric redshift technique is set out.  In \S\ref{sec:seds} the various templates and extensions to the UV are discussed.  The ImpZ code is then applied to two spectroscopic redshift galaxy samples in \S\ref{sec:specz}.  The results of this validation are given in \S\ref{sec:specresults}.  Error analysis is discussed in \S\ref{sec:error}.  Discussions and conclusions are presented in \S\ref{sec:disc} and \S\ref{sec:conc}.

The application of the ImpZ code to the entire re-calibrated ELAIS N1 and N2 fields from the INT WAS and investigations into the evolution of extinction and star formation rates (SFR) will be presented in a companion to this paper; Babbedge et al. (2004; in prep.).\\
%Note that for these investigations the canonical Einstein de--Sitter model with H$_0$=65km s$^{-1}$Mpc$^{-1}$ is used. 
Note that for these investigations the flat, $\Omega_\Lambda$=0.7 cosmological model with H$_0$=65km s$^{-1}$Mpc$^{-1}$ is used. 
%%%%%%%%%%%%%
\begin{table*}
\caption{\scriptsize{Common spectral lines and breaks in SEDs, and the redshifts where they enter and leave the filters used in this study, where transition is defined by the points where filter transmission drops below 50 per cent of the maximum filter transmission.  Redshifts greater than seven are not considered.}}
\begin{tabular}{|c c c c c c c c c c c|}
\hline
\hline
\multicolumn{11}{c}{$INT$ $WAS$ $filters$}\\
\hline
$Filter$&\multicolumn{2}{c}{Lyman Limit ($912{\rm \AA}$)}&\multicolumn{2}{c}{Ly$\alpha$ $(1216{\rm \AA}$)}&\multicolumn{2}{c}{Balmer Break ($4000{\rm \AA}$)}&\multicolumn{2}{c}{$\overline{0[III]}$ doublet ($4983{\rm \AA}$)}&\multicolumn{2}{c}{H$\alpha$ ($6563{\rm \AA}$)}\\
\cline{1-11}
& enters & leaves& enters & leaves& enters & leaves& enters & leaves& enters & leaves \\
$U$&2.5&3.3&1.6&2.2&--&--&--&--&--&--\\
$g\arcmin$&3.5&4.9&2.4&3.4&0.03&0.35&0&0.08&--&--\\
$r\arcmin$&5.0&6.6&3.5&4.7&0.38&0.73&0.10&0.38&005&0.07\\
$i\arcmin$&6.7&NA&4.8&6.0&0.75&1.1&0.40&0.71&0.07&0.30\\
$Z$&NA&NA&6.0&6.9&1.1&1.4&0.71&0.93&0.30&0.46\\
\hline
\hline
\multicolumn{11}{c}{$Barger$ $et$ $al.$ $(2000)$ $filters$}\\
\hline
$Filter$&\multicolumn{2}{c}{Lyman Limit ($912{\rm \AA}$)}&\multicolumn{2}{c}{Ly$\alpha$ $(1216{\rm \AA}$)}&\multicolumn{2}{c}{Balmer Break ($4000{\rm \AA}$)}&\multicolumn{2}{c}{$\overline{0[III]}$ doublet ($4983{\rm \AA}$)}&\multicolumn{2}{c}{H$\alpha$ ($6563{\rm \AA}$)}\\
\cline{1-11}
& enters & leaves& enters & leaves& enters & leaves& enters & leaves& enters & leaves\\
$B$&3.3&4.5&2.2&3.1&0&0.25&0&0.003&--&--\\
$V$&4.5&5.6&3.1&3.9&0.25&0.50&0003&0.20&--&--\\
$R$&5.6&6.7&3.9&4.8&0.50&0.75&0.20&0.40&0&0.07\\
$I$&6.9&NA&4.9&6.1&0.80&1.2&0.44&0.73&0.10&0.31\\
$Z$&NA&NA&5.5&NA&1.1&1.5&0.69&1.0&0.28&0.52\\
$HK\arcmin$, ($H$)&NA&NA&NA&NA&2.7&3.5&2.0&2.6&1.2&1.7\\
$HK\arcmin$, ($K\arcmin$)&NA&NA&NA&NA&3.9&4.8&2.9&3.6&2.0&2.5\\
\hline
\hline
\end{tabular}
\label{table:features}
\end{table*}
%%%%%%%%%%%%%
%%%%%%%%%%%%%%%%%%%%%%%%%%%%%%%%%%%%%%%%%%%%%%%%%%%%%%%%%%
%%%%%%%%%%%%%%%%%%%%%%%%%%%%%%%%%%%%%%%%%%%%%%%%
\section{Method}
\label{sec:method}
%%%%%%%%%%%%
\subsection{$\chi^2$ analysis}
\label{subsec:chi}
The template-fitting procedure is as follows:  the observed galaxy magnitudes are converted for each $i^{th}$ photometric band into an apparent flux, $f_{i}^{obs}$.  Equivalently, this reconstructs the target galaxy's SED at a very low spectral resolution by sampling the luminosity at the effective wavelength of each photometric band.  The observed fluxes can then be compared to the template fluxes, $f_{i}^{templ}$(z,T), computing the reduced $\chi^{2}$, $\chi^{2}_{red}$ as
\begin{equation}
\label{eqn:chi}
\chi^{2}_{red}=1/D\sum_{i=1}^{N}\frac{[f_{i}^{obs}-sf_{i}^{templ}(z,T)]^2}{\sigma_{i}^2}
\end{equation}
where N is the number of photometric bands, D is the number of degrees of freedom, $\sigma_i$ is the observational uncertainty in the $i^{th}$ band (hence the solution is weighted by the flux errors, however to avoid excessively high weighting by very high signal-to-noise observations, a minimum flux error is set, typically 0.5 per cent) and s is a normalisation factor to minimize $\chi^{2}$ for each template.%and that in this paper, rather than using a set of model fluxes spaced at regular intervals in redshift space, log$_{10}$(1+z) space is used so that model fluxes are calculated at intervals of 0.01 in log$_{10}$(1+z).  This is more efficient since it samples lower redshifts to a high resolution whilst not needlessly oversampling higher redshifts.
\begin{equation}
\label{eqn:s}
s=\frac{\sum_{i=1}^{N}f_{i}^{obs}f_{i}^{templ}(z,T)/\sigma_{i}^2}{\sum_{i=1}^{N}f_{i}^{templ}(z,T)^2/\sigma_{i}^2}
\end{equation}
%It is this photometric method which will be developed and applied to the two spectroscopic samples.  
Note that if there is a detection in just one band then fitting is not attempted.
%%%%%%%%%%%%
\subsection{The ImpZ code}
\label{subsec:code}
\begin{figure*}
\begin{center}
\includegraphics[width=14cm,height=5.5cm]{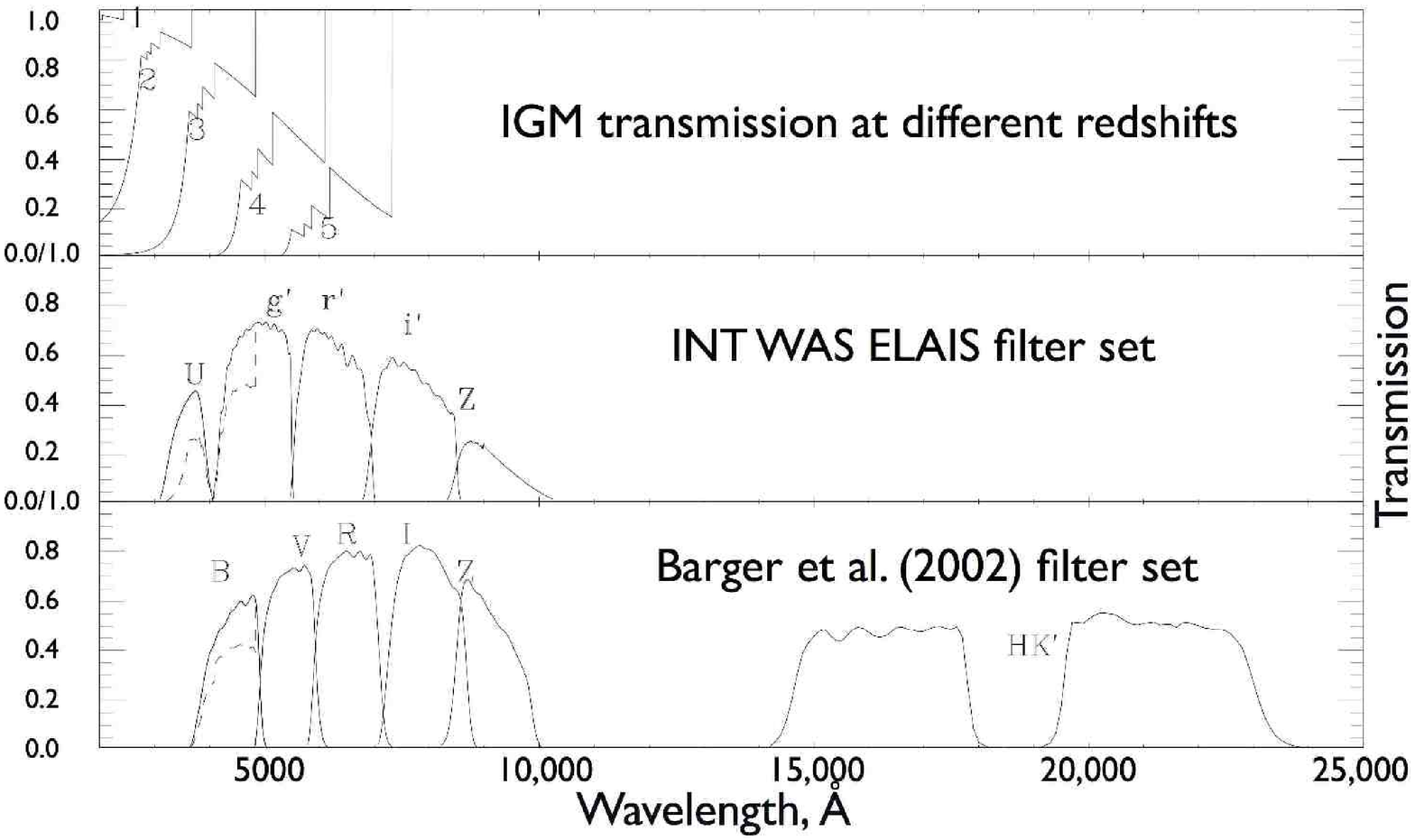}
\caption{\scriptsize{IGM transmission effect at redshift 1, 2, 3, 4 and 5 on the observed flux from a source.  Below are shown the INT WAS filters (solid lines), combined with the CCD response of the WFS camera, and also the filters (solid lines) used in the Barger et al. (2002) CDFN catalogue, combined with the relevant CCD (Suprime--Cam and QUIRC) responses.  The additional effect of including IGM absorption is illustrated for the redshift 3 case by over-plotting the filter response when combined with the redshift 3 IGM transmission curve.  It can be seen to effect $U$ and $g\arcmin$ in the INT WAS filter set, and the $B$ filter in the CDFN filter set (over-plotted as dashed lines).}}\label{fig:igmfilters}
\end{center}
\end{figure*}

This code builds on the technique presented in RR03, extending it to include the correct treatment of CCD response, filter transmission characteristics and the statistical effect of Inter-Galactic Medium (IGM) absorption (Ly$\alpha$, Ly$\beta$, Ly$\gamma$, Ly$\delta$, and Lyman continuum),  as set out in \cite{Madau1996MNRAS.283.1388M} (see Fig. \ref{fig:igmfilters}).  Correct treatment of the effects of the IGM is needed because it is possible to mis-interpret the Ly$\alpha$ forest effect as the intrinsic Lyman break and because, particularly at high redshifts, it will imprint its own recognisable feature onto the SED.  Although the treatment of IGM absorption is only based on the average accumulated absorption, the $rms$ fluctuations away from the mean can be expected to be small once integrated though a broad bandpass \citep*{Press1993ApJ...414...64P}.  At high redshift, \cite*{Massarotti2001A&A...368...74M} have shown that correct treatment of internal dust reddening (the Interstellar Medium) and IGM attenuation are the main factors in photometric redshift success.  The effect of internal dust reddening for each galaxy is already incorporated in the templates (Table \ref{fig:seds}) and alterable via fitting for A$_{v}$, using the reddening curve of \cite{Savage1979ARA&A..17...73S}.  Observed fluxes are compared to template fluxes for 0.01$\leqslant$log$_{10}$(1+z)$\leqslant$0.90, equivalent to 0.02$<$z$\leqslant$6.94.

The following parameters and cuts were used:
\begin{enumerate}
\item in order to reject unphysical fits only those that give absolute $B$--band AB magnitudes (M$_B$, 4400{\rm \AA}) in the range [$-22.5-2log_{10}(1+z)]<$M$_B<-13.5$ are considered for the 6 galaxy templates.  Having an upper envelope dependent on redshift was found to be the best way to suppress luminous outliers at low redshift whilst allowing more luminous galaxies at higher redshifts, and this
luminosity--redshift dependence agrees with the natural consequence of known strong
luminosity evolution for galaxies (e.g. \citealt{Lin1999ApJ...518..533L}).  For the AGN templates a range from $-27<$M$_B<-17.5$ was allowed.  These restrictions cut out excessive numbers of aliases at the minimum and maximum redshift.  A number of different limits have been investigated but these were found to be the most effective.  Originally, limits were applied to the $I$ band but this failed to constrain the UV luminosity.  Shifting the limits to the $B$ band allows both the young and old star components of the SED to be constrained for the redshifts considered.
%Dropouts were treated in the following manner: If there is an upper limit in U or g$\arcmin$ which is more than a factor 4 below the flux of the next shortest wavelength band then this upper limit is used in the photometric solution.\\
\item Sources are defined as $stellar$ or $non$-$stellar$, where this is a purely morphological property differentiating between point-like and more extended sources.  Those defined as $stellar$ will have Active Galactic Nuclei (AGN) templates considered in addition to the galaxy templates, whereas $non$-$stellar$ sources will only be fit by galaxy templates.  This reduces the increased degeneracy introduced by including AGN templates.  The procedure for splitting sources into $stellar$/$non$-$stellar$ sub-groups is different for the INT WAS ELAIS spectroscopic sample and the CDFN sample, and is described for each in \S\ref{subsec:wfsspeccat} and \S\ref{subsec:cdfnspeccat} respectively.
\item sources that were saturated were removed since their photometry is poor.  This reduces z$_{phot}$ outliers.
\item A prior expectation that a $stellar$ source is more likely to be a QSO is introduced by minimising $\chi^{2}_{red}+\alpha F(T)$ rather than $\chi^2$ ($\alpha$=4 here) for $stellar$ sources.  $F(T)$ is a delta function such that $F(T)=0$ if the template, $T$, is an AGN template, and $F(T)=1$ if the template is a galaxy template.  Essentially it is a prior to prefer AGN solutions due to the morphology information contained in the class flags -- a weak Bayesian (e.g. \citealt{Jaynes2000}) formulation -- and was reached based on the results in \S\ref{subsubsec:relaxation} and \S\ref{subsubsec:cdfnrelaxation}.

For the `free' A$_{v}$ fitting option the following restrictions were used:
\item for the elliptical and AGN templates, A$_{v}$ can take the value 0 only.  The reason that A$_{v}$ is set to zero for AGN is that since AGN are essentially a power-law, the additional inclusion of A$_{v}$ gives too much freedom to the shape of the AGN template, and it was found that resulting degeneracies  reduce the effectiveness of the photometric redshift technique.%  An attempt to include reddened AGN is instead carried out by the inclusion of a red AGN template (see \S\ref{subsec:agn}). 
\item for other templates, A$_{v}$ can take the values 0.0 to 1.0, in steps of 0.1.  The maximum A$_{v}$ was chosen to be approximately twice that of the typical A$_{v}$ of galaxies at $z\approx4$ found by \cite{Steidel1999ApJ...519....1S} who derived a typical $E(B-V)$ of 0.15.  Note that the templates already include some extinction (see Table \ref{table:seds}) so that the A$_{v}$ of the solution is technically the difference between the actual value and the template's inherent value.%  \cite{King2003MNRAS.339..260K} found that the mean local Universe A$_{v}$ of their best-fitting source-count model was roughly 0.4.
\item no solution for A$_{v}$ is sought if the reduced $\chi^{2}$, $\chi^{2}_{red}$, of the A$_{v}$=0 solution is $<$1, or if there are less than 4 bands.
\item a prior expectation that the probability of a given value of A$_{v}$ declines as $|$A$_v|$ moves away from 0 is introduced by minimising $\chi^{2}_{red}$ + $\beta$A$_v^2$ rather than $\chi^2$ ($\beta$=2 here).  This can again be viewed as a weak implementation of Bayesian methods.
\end{enumerate}
%%%%%%%%%%%%%%%%%%%%%%%%%%%%%%%%%%%%%%%%%%%%%%%%
%%%%%%%%%%%%%%%%%%%%%%%
%%%%%%%%%
\section{Template SEDs}
\label{sec:seds}
The choice of how many templates to use in fitting is a crucial one.  The choice of too many leaves the code with too much freedom, leading to large numbers of aliases and degeneracy.  Similarly, too few and the code will be unable to find accurate redshifts for real objects, something \cite{Bolzonella2000A&A...363..476B} termed `catastrophic' failures.

As in \cite{Mobasher1996MNRAS.282L...7M} and RR03 the code uses six galaxy templates; E, Sab, Sbc, Scd, Sdm and starburst galaxies.  These six templates (or similar versions) have been found to provide a good low-resolution representation of observed galaxy SEDs (\citealt{Mobasher1996MNRAS.282L...7M}, RR03), and are used in preference to a large array of evolving galaxy templates, as generated by evolutionary codes such as that of \cite{Bruzual1993ApJ...405..538B}, since too much freedom is then available in fitting.  The empirical templates used in RR03 were based on observations (starburst template adapted from \citealt{Calzetti1992ApJ...399L..39C}) and on observations and colour synthesis (the remainders adapted from \citealt{Yoshii1988ApJ...326....1Y}).  In a method reminiscent of empirical techniques, these templates were adapted to improve photometric redshift results.
%  Instead, evolution is expected to be seen in the proportion of different templates that are fit in different redshift bins.

\begin{figure*}
\begin{center}
\includegraphics[width=11cm,height=15cm,angle=90]{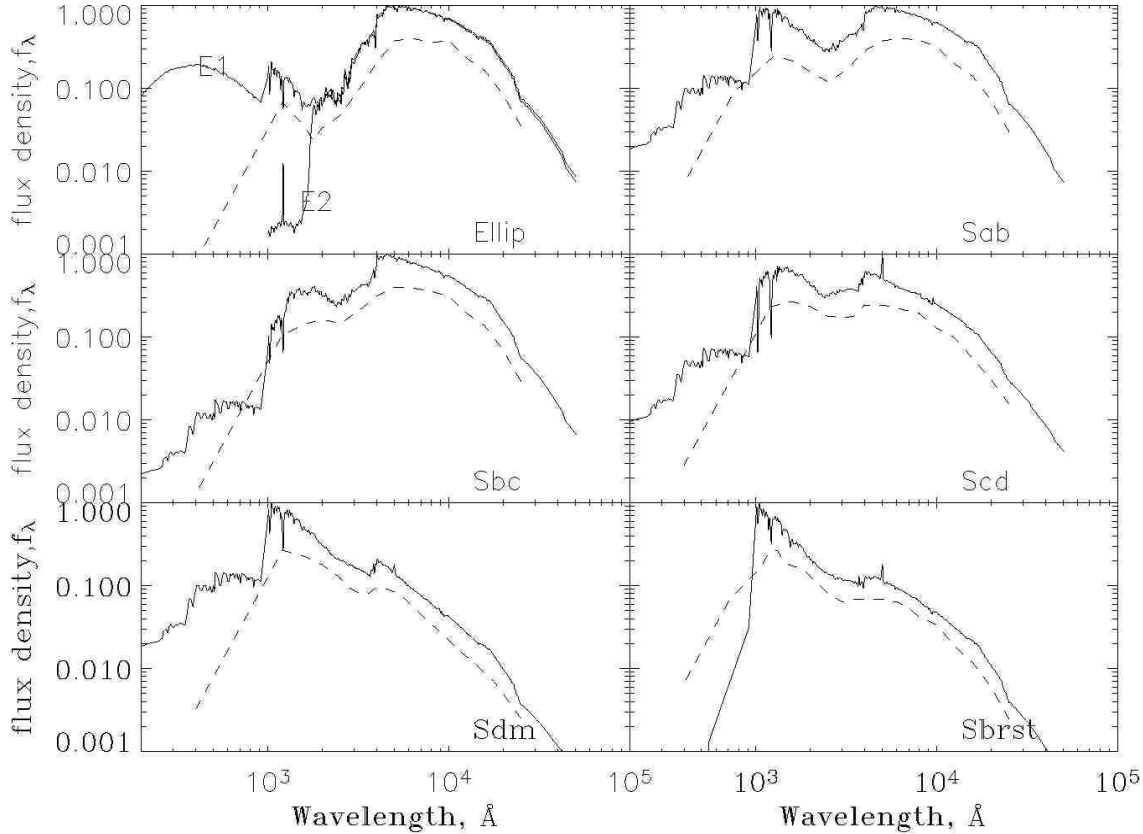}
\caption{\scriptsize{The 6 galaxy templates used.  Dashed lines show the original RR03 templates (offset for clarity), solid lines shows the SSP generated versions, along with extension into the Far-UV (sub--1000{\rm \AA}) as set out in section \ref{subsec:uv}.  For the elliptical template two SSP generated fits are shown, which diverge below around 2000{\rm \AA}.  Line E1. fits the UV bump that is due to planetary nebulae, whereas Line E2. does not.  This is discussed in section \ref{subsec:ssps}.}}\label{fig:seds}
\end{center}
\end{figure*}

The templates used in this work were generated by reproducing these original templates via spectrophotometric synthesis, in order to strengthen their physical basis.  The original templates were convolved with filter transmission curves in order to create virtual datapoints.  These datapoints were then fit using a code which combines a given number of  Simple Stellar Populations (SSPs), each weighted by a different SFR and extinguished by a different amount of dust. The procedure is based on the synthesis code of \cite*{Poggianti2001ApJ...550..195P} and has been previously applied to spectra in \cite{Berta2003A&A...403..119B}.  The code minimizes the $\chi^2$ obtained from comparing the new template to the datapoints. Minimization is based on the Adaptive Simulated Annealing algorithm.  Details on this algorithm and on the fitting technique are given in \cite{Berta2004A&A...418..913B}.  The templates are plotted in Fig. \ref{fig:seds}, and have been extended further into the UV regime as set out in \S\ref{subsec:uv}.
%%%%%%%%%%%%%%%%%%%%%%%%%%%%%%%%%%%%%%%%%%%%%%%%
%%%%%%%%%%%%%%%%%%%%%%%
\subsection{SSP Populations}
\label{subsec:ssps}
The spectra of the SSPs have been computed with a Salpeter Initial Mass Function (IMF) between 0.15 and 120 solar masses, adopting the \cite{Pickles1998PASP..110..863P} spectral atlas and extending its atmospheres outside its original range of wavelengths with \cite{Kurucz1993KurCD..13.....K} models from 1000{\rm \AA} to 50,000{\rm \AA}, as described in \cite*{Bressan1998A&A...332..135B}.  Nebular emission is added by means of case B HII region models computed through the ionization code CLOUDY of \cite{Ferland19981998PASP..110..761F}. The adopted metallicity is solar.

When generating the new templates, each SSP is weighted by a different SFR. For each SSP a uniform screen attenuation is adopted, using the standard extinction law (Rv=A$_{v}$/($E(B-V)$=3.1, \citealt*{Cardelli1989ApJ...345..245C}) and adopting a different $E(B-V)$.

With 10 SSPs, we have a total of 20 free parameters (SFR and $E(B-V)$ for each SSP), but the code automatically discards those populations which contribute less than one per cent to the total spectrum, at each wavelength.  As a result each template is constructed from only a few SSPs.  The $E(B-V)$ values for the oldest populations (9th and 10th) are constrained to be less than 0.2 (the characteristic extinction of the older quiescent stellar population in a sample of nearby galaxies is $E(B-V)$$\approx0.1$; \citealt{RR2003MNRAS.344...13R}).  The total $E(B-V)$ for each template is obtained by comparing the non-extincted final spectrum $(S_n)$ to the extincted version $(S_e)$ since in the $V$ band the following relation holds:  $S_e=10^{-0.4A_v}S_n$.  For details of the populations and their contribution to each template see Table \ref{table:seds}.  

 %Without this stipulation, it was found that the oldest populations had excessive reddenning, $E(B-V)$$>$0.3, whereas the characteristic extinction of the older quiescent stellar population in a sample of nearby galaxies is $E(B-V)$$\sim$0.1 (\citealt{RR2003UVastroph}).  For details of the populations and their contribution to each template SED see Table \ref{table:seds}.\\
Two fits were generated for the elliptical template $-$ E1 and E2.  E1 fits the small UV bump present in ellipticals from approximately 1000--2000{\rm \AA}, a feature that is due to emission from planetary nebulae (\citealt{Yoshii1988ApJ...326....1Y}).  In order to prevent this UV bump being fit by young stellar populations, the SFR of the 3 youngest SSPs was set to zero.  E2 consists only of the two oldest SSPs and fails to fit this UV bump.  For this reason, E2 was not used as a template, but is plotted in Fig. \ref{fig:seds} for interest.

The template--extension into the UV is described in \ref{subsec:uv}.
\begin{table*}%[!ht]
\caption{\scriptsize{The six galaxy templates and the SSPs that were used to create them.  SFR is given in M$_{\odot}yr^{-1}$ and normalised so that each template has a total mass of 10$^{10}$M$_{\odot}$.}}
\centering
\scriptsize
\begin{tabular}{c r r | c*{12}{c}}
\hline
\hline
\multicolumn{3}{c|}{SSP} 	&\multicolumn{12}{c}{SED Template}\\
\hline
\#& \multicolumn{1}{c}{Age (yr)} & \multicolumn{1}{c|}{$\Delta t$ (yr)} &&\multicolumn{2}{c}{Elliptical(E1)}&\multicolumn{2}{c}{Sab}&\multicolumn{2}{c}{Sbc}&\multicolumn{2}{c}{Scd}&\multicolumn{2}{c}{Sdm}&\multicolumn{2}{c}{Sbrst}\\
\cline{5-16}
%\cline{7-8}
%\cline{8-9}
%\cline{14-15}
%\cline{17-18}
 %\cline{20-21}
    	&		&	       && SFR  & E(B-V) &SFR  & E(B-V)  & SFR     & E(B-V) & SFR   & E(B-V) &SFR  & E(B-V) & SFR &E(B-V) \\
1	&$10^6$		&$2\, 10^6$     && -- &   --  & -- &  -- & --   & -- &  --  & -- & -- & -- & -- & -- \\
2	&$3\, 10^6$	&$4\, 10^6$     && -- &  -- &  0.100 &   0.04  & --  &  -- &  --  &--& 4.097 &  0.001  & 41.17 &0.9465\\
3	&$8\, 10^6$	&$3\, 10^6$     &&  --  &   --  &   --  &   --  &  --     &	--  &  --   &  --  &  --  &  --  & 35.67  &  0.024  \\
4	&$10^7$		&$4\, 10^6$     && 0.099 &   --  &   --  &   --  & 1.461     &	0.18  & --   &  --  & --  &  --  & --  &  --  \\
5	&$5\, 10^7$	&$6.2\, 10^7$   &&  --  &   --  &0.332  &   0.015  & --     &	--  & 2.576  &  0.03475  &13.54 & 0.004  & --  &  --  \\
6	&$10^8$		&$1.25\, 10^8$  &&  --  &   --  &  --  &   --  & --     &	--  &  --  & -- &  13.62  &  4.685  & --  &  --  \\
7	&$3\, 10^8$	&$2\, 10^8$     &&-- &  -- & 0.484 &  0.12 &1.647     &	--  &1.692   &  0.12 &9.995 &  --  & --  &  --  \\
8	&$5\, 10^8$	&$3.5\, 10^8$   && -- &  -- & -- &  -- & 0.896   &  0.003 &2.80  & 0.032 &11.58 & 0.0616 &  --  &  --  \\
9	&$10^9$		&$1.25\, 10^9$  &&  0.396  &   --  &1.074  &   0.017  &1.509   &  0.137 &0.771   &  0.1838  &1.111  &  0.0818  &7.78 & 0.0015 \\
10	&$1.2\, 10^{10}$&$10^{10}$      && 0.950 &  -- &0.854 &  0.134 & 0.747    &  0.186 &0.756 & 0.138 &-- & -- &--& --  \\
\hline
%\multicolumn{3}{l|}{M$_{tot}$}&&\multicolumn{2}{c}{5.69e9}&& \multicolumn{2}{c}{1.05e10} && \multicolumn{2}{c}{5.97e10} && \multicolumn{2}{c}{1.77e10} && \multicolumn{2}{c}{1.23e10} && \multicolumn{2}{c}{5.12e9}\\
%\hline
\multicolumn{3}{l|}{$A_v$(young - 1st 3 SSPs)}
&&\multicolumn{2}{c}{0.0}&\multicolumn{2}{c}{0.12205}&\multicolumn{2}{c}{0.0}&\multicolumn{2}{c}{0.0}&\multicolumn{2}{c}{0.0031}&\multicolumn{2}{c}{1.38}\\
\multicolumn{3}{l|}{$A_v$(total)}
&&\multicolumn{2}{c}{0.0}&\multicolumn{2}{c}{0.2344}&\multicolumn{2}{c}{0.3308}&\multicolumn{2}{c}{0.271}&\multicolumn{2}{c}{0.4303}&\multicolumn{2}{c}{0.736}\\
\hline
\end{tabular}
\footnotesize
\label{table:seds}
\normalsize
\end{table*}
%%%%%%%%%%%%%%%%%%%%%%%%%%%%%%%%%%%%%%%%%%%%%%%%%%%%%%%%%%
\subsection{AGN Templates}
\label{subsec:agn}
\begin{figure}
%\begin{center}
\includegraphics[width=5.7cm,angle=90]{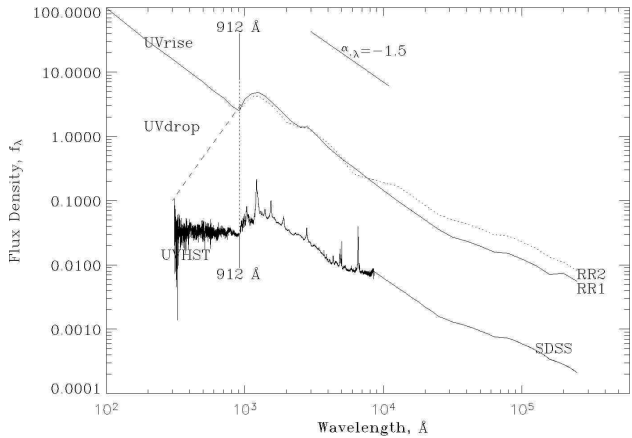}
\caption{\scriptsize{The various AGN templates that were investigated.  The solid line `SDSS' is the mean SDSS quasar spectra (Vanden Berk et al. 2001), shown here with the Zheng et al. (1997) UV behaviour, `UVHST', and extended into the IR as set out in \S\ref{subsec:agn}.  The solid line `RR1' and dotted line `RR2' are the empirical AGN templates based on Rowan--Robinson (1995), shown with either a drop-off in the UV (`UVdrop', dashed) or a rise in the UV (`UVrise', solid).  The 912{\rm \AA} Lyman limit has been indicated, as is the slope of a power-law continuum with $\alpha_{\lambda}$=$-$1.5.  Discussion of these AGN is covered in \S\ref{subsec:agn}, with UV treatment discussed in \S\ref{subsec:uv}.}}\label{fig:agn}
\end{figure}

As well as galaxy templates, the inclusion of a number of different AGN templates has been investigated (see Fig. \ref{fig:agn}) to allow the ImpZ code to identify quasar-type objects as well as normal galaxies.  This is of particular interest for application of ImpZ to the entire ELAIS N1 and N2 fields of the INT WAS in Babbedge et al. (2003; in prep.) since many ELAIS sources are expected to be AGN (\citealt{Oliver2000MNRAS.316..749O}).  Fitting with AGN templates is only carried out for sources that have been defined as $stellar$, as described in \S\ref{subsec:code}.

The last decade has seen a large rise in the number of optically selected high-redshift quasars and the existence of large samples of quasars (e.g the Sloan Digital Sky Survey -- SDDS -- \citealt{York2000AJ....120.1579Y}) means that the derivation of photometric redshifts for quasars is gaining popularity as well as reliability.  For galaxies the technique relies on the identification of continuum features such as the 4000{\rm \AA} break -- see Table \ref{table:features} for examples.  For a featureless spectra a photometric redshift is far harder to determine, if at all.  The majority of quasars can be characterised in the UV--optical region as a featureless continuum.  Overlaid on this continuum however are a series of (mostly) broad emission line features which contain a significant amount of flux (\citealt{Francis1991ApJ...373..465F}; \citealt{Richards2001AJ....121.2308R}).  Also, at higher redshifts the result of Ly$\alpha$ forest absorption will imprint an additional redshift-dependent feature onto the continuum.  Empirical redshift--colour relationships have been applied to quasars, but such polynomial fitting is limited due to the nature of quasar colours, which can change rapidly across a small redshift range as an emission line passes in and out of a passband, or remain constant with redshift due to the featureless continuum -- a relationship which is poorly reproduced by polynomial functions that vary slowly with redshift.  An extension of this technique which implements a nearest neighbour (NN) estimator with the reference points derived from colours averaged over redshift bins was presented in \cite{Richards2001AJ....122.1151R}, with around 70 per cent of predicted redshifts matching reasonably well with the spectroscopic values ($\Delta z<0.1$ for 55 per cent and $\Delta z<0.2$ for 70 per cent of their sample).

The alternative approach is a template-fitting procedure as used in several quasar studies (e.g. \citealt*{Hatziminaoglou2000A&A...359....9H}) and adopted in this work.  This approach requires a template or templates that can cover the observed range of spectral types, across a large enough wavelength span to be applicable for low and high redshift objects.  Unlike galaxies, quasars have SEDs with similar power-law continua so the use of the mean spectrum of a sample of quasars is feasible.  \cite{Budavari2001AJ....122.1163B} used the SDSS composite spectrum of \cite{VandenBerk2001AJ....122..549V}, resulting in a slightly greater $rms$ than that found in the empirical NN method of \cite{Richards2001AJ....122.1151R}.  This suggests that the use of a single quasar template is not sufficient -- perhaps one corresponding to broad absorption lines or FeII is also needed.  A similar study by \cite*{Wu2004ChJAA...4...17W} with the SDDS composite achieved $\Delta z<0.1$ for 47 per cent and $\Delta z<0.2$ for 68 per cent of their sample.  \cite{Budavari2001AJ....122.1163B} went on to reconstruct 4 discrete templates in an iterative manner, gaining better results than with the empirical NN method.   However, this is beyond the scope of this paper and is left as a future direction.\\

In this study, several AGN templates were constructed.  The optical--IR basis of these templates is outlined below, and extension to UV  in \S\ref{subsec:uv}:\\
\begin{enumerate}
\item the SDSS median composite quasar spectrum (\citealt{VandenBerk2001AJ....122..549V}) covers a rest-wavelength range from 800 to 8555{\rm \AA} and is constructed from 2204 quasars spanning 0.044$\leqslant z\leqslant$4.789.  For wavelengths longward of Ly$\alpha$ to optical wavelengths the continuum is well-fitted by a power law (f($\lambda)\propto\lambda^{\alpha_{\lambda}}$) with a wavelength power-law index, $\alpha_{\lambda}$ ($\alpha_{\lambda}$=$-$($\alpha_{\nu}$+2)), of $-$1.56.  This is consistent with a number of other works based on optical and/or radio-selected samples that have found power-law continuum indexes, as average values from spectra, from photometry, or using composite spectra, of $-2\leqslant\alpha_{\lambda}\leqslant-1$ (e.g. \citealt{Schneider2001AJ....121.1232S}, \citealt{Brotherton2001ApJ...546..775B}, \citealt{Carballo1999MNRAS.306..137C}, \citealt{Natali1998AJ....115..397N}, \citealt{Zheng1997ApJ...475..469Z}, \citealt{Francis1996PASA...13..212F}, \citealt{Francis1991ApJ...373..465F} and \citealt{Cristiani1990A&A...227..385C}).  The continuum blueward of the Ly$\alpha$ emission line is heavily absorbed (the median redshift is 1.253) due to Ly$\alpha$ forest absorption.  Since this effect is strongly effected by redshift however and the median composite uses spectra across a broad range of redshift, little can be drawn from the absorption in that region of the SED.  Instead, the UV part of the SED will be treated separately -- see \S\ref{subsec:uv}.  The template has been extended to wavelengths longer than 8555{\rm \AA} by utilising the IR part of the average optical quasar  spectrum of \cite{RR1995MNRAS.272..737R}, slightly modified as set out in \cite{RR2004MNRAS.345..1290R}.  This extends the template out to 25$\mu$m with what is essentially a continuation of the continuum power-law.  See the `SDSS' line in Fig. \ref{fig:agn}.
\item in addition to the SDSS composite template, two simpler AGN templates were included.  These are based on the mean optical quasar  spectrum of \cite{RR1995MNRAS.272..737R}, spanning 400${\rm \AA}$ to 25$\mu$m.  For wavelengths longer than Ly$\alpha$ the templates are essentially $\alpha_{\lambda}\approx-1.5$ power-laws, with slight variations included to take account of observed SEDs of ELAIS AGN \citep{RR2004MNRAS.345..1290R}.  These two AGN templates are referred to as RR1 and RR2, and are very similar to those used in \cite{RR2004MNRAS.345..1290R}.  They differ from one another at wavelengths longer than 1$\mu$m where RR2 contains more flux.  For wavelengths shorter than the Lyman limit, several UV behaviours were again considered -- see \S\ref{subsec:uv}.  These templates can be seen as `RR1' and `RR2'  in Fig. \ref{fig:agn}.
\end{enumerate}
\subsubsection{A $red$ quasar}
\label{subsubsec:redqso}
Should a much redder quasar be considered?  There is a debate over the existence of a significant population of $red$ quasars (e.g. \citealt{Webster1995Natur.375..469W} and \citealt{Brotherton2001ApJ...546..775B}).  \cite{Richards2003AJ....126.1131R} found that roughly 6 per cent of their homogeneously selected sample of 4576 SDSS quasars were red in comparison with even the reddest power-law continuum quasars and were probably dust reddened.  In this work a $red$ AGN template has not been included -- some tests were done using the $z=2.216$ FIRST J013435.7--093102 source from \cite{Gregg2002ApJ...564..133G}, which is an extremely dust--reddened lensed object with B$-$K$\geqslant10$ but it was found that it did not match any sources in our two spectroscopic redshift catalogues.  The inclusion of a red AGN template is, however, expected to be powerful when optical data is combined with upcoming IR data from the $SPITZER$ mission, particularly for the $SPITZER$ Wide--Area Infrared Extragalactic Survey (SWIRE; \citealt{Lonsdale2003PASP..115..897L}) which covers an area large enough (50 square degrees) to find a significant number of these rare objects.  Application to SWIRE data will be considered in a future work (Babbedge 2004; PhD thesis, in prep.).
%%%%%%%%%%%%%%%%%%%%%%%%%%%%%%%%%%%%%%%%%%%%%%%%
%%%%%%%%%%%%%%%%%%%%%%%
\subsection{Far UV treatment}
\label{subsec:uv}
The Milky Way becomes virtually opaque at wavelengths between 100 -- 912{\rm \AA} due to absorption by neutral hydrogen.  Hence it is difficult to be certain of the far-UV rest-frame emission of galaxies or quasars.  Ground-based observations can only start revealing the sub--1200{\rm \AA} regime for objects at $z\ga$2 due to atmospheric absorption of shorter wavelengths.  Escaping our own atmosphere with space-based telescopes has allowed the far-UV and extreme-UV region to be explored in more detail, but even then the action of the IGM and of neutral hydrogen in our own Galaxy makes the SED determination uncertain.  In order to apply a template-fitting technique out to large redshifts it is important that the templates extend to sufficiently short wavelengths.  For galaxies this is perhaps a more straight-forward task, since the emitted electromagnetic energy can be assumed to be due to stars and dust heated by those stars.  There exist a number of stellar synthesis codes that, when coupled with spectral evolution, can self-consistently reproduce the emission of galaxies across many orders of magnitude in wavelength.  The SSPs adopted by the spectral synthesis code and used to generate the 6 galaxy templates in this work were not detailed enough below 1000{\rm \AA};  hence for the far--UV part of these templates the results of other spectral evolution codes has been considered.

Both the isochrone synthesis of \cite{Bruzual1993ApJ...405..538B} and the PEGASE code of \cite{Fioc1997A&A...326..950F} show that for old stellar populations (older than several Gyr) such as those that characterise ellipticals, there is a rise in the far-UV due to low-mass stars in their post-AGB evolution.  Furthermore, hot post-AGB stars decrease the amplitude of the 912{\rm \AA} break once their envelopes have dissipated.

In order to extend the elliptical template into the far-UV, therefore, the sub--912{\rm \AA} part of the spectrum from the elliptical template in HYPERZ (\citealt{Bolzonella2000A&A...363..476B}) was used, who had extended it from the elliptical template of \cite*{Coleman1980ApJS...43..393C}, hereafter CWW.  The flux was scaled in order to give a slight rise across the 912{\rm \AA} discontinuity.  The four template set of CWW (elliptical, Sbc, Scd, Im) has been used in many photometric redshift studies (E.g \citealt{Gwyn1999prdh.conf...61G}, \citealt{Benitez1999prdh.conf...31B} and \citealt{Brodwin1999prdh.conf..105B}) and has been found to be a robust and reasonably complete template set.  The extension into the far-UV carried out by \cite{Bolzonella2000A&A...363..476B} was by means of \cite{Bruzual1993ApJ...405..538B} spectra with parameters (SFR and age) selected to match the observed spectra at zero redshift.  They infact use the IMF of \cite{Miller1979ApJS...41..513}, however this choice has a negligible impact on the final results, as they discuss in their \S 4.6.

The Sab, Sbc, Scd and Sdm templates were extended in a similar manner, using the sub--912{\rm \AA} part of the spectrum (suitably scaled to give a factor 10 rise in flux across the 912{\rm \AA} discontinuity) from the Scd template of CWW, extended by \cite{Bolzonella2000A&A...363..476B}.  They were all given the same UV behaviour because the difference in UV spectrum between these galaxy types is expected to be less than the uncertainty in their actual UV behaviour.  Indeed at redshifts where this region enters the optical filters, the dominant effect is due to IGM absorption.\\
The starburst template was extended following the results of  \cite{Bruzual1993ApJ...405..538B} and \cite{Fioc1997A&A...326..950F} which show that within 10Myr the UV light drops sharply due to the evolution of massive stars off the Main Sequence.  Below 912{\rm \AA} the starburst was assumed to be optically thick, an assumption that has been verified for nearby starburst galaxies by \cite{Leitherer1995ApJ...454L..19L}.

It is noted here that as well as the UV behaviours used above for the 6 galaxy templates, other forms were tested, such as sharp cut-offs at 912{\rm \AA} -- essentially assuming that the galaxies are optically thick to ionizing radiation below the Lyman limit -- or simply taking the flux at 1000{\rm \AA} and setting this value for sub--1000{\rm \AA} wavelengths.  Such approaches were not found to be as successful.\\

Determining the exact shape of the UV spectra of AGN is problematic for the same reasons as for galaxies.  Additionally, the observed broad continuum feature in the optical--UV, the `big blue bump', is confused by many broad and blended lines, which are thought to be due to fast moving ionised clouds near the centre of the AGN.  Contamination from the host galaxy also has an effect for low luminosity AGN, particularly in the case of a nuclear starburst.  Hence several far-UV trends have been investigated to see how they effect the accuracy of quasar photometric redshifts:
\begin{enumerate}
\item although observations of the continuum blueward of the Lyman edge are rare, \cite{Zheng1997ApJ...475..469Z} constructed a composite spectrum from 284 $Hubble$ $Space$ $Telescope$ (HST) Faint Object Spectrograph (FOS) spectra of 101 quasars at $z\approx1$.  Around 90 per cent of the sample were at redshifts $<$ 1.5 so the region blueward of Ly$\alpha$ could be studied without large effects from the Ly$\alpha$ forest.  The shortest wavelength data, 350--600{\rm \AA}, were drawn from higher redshift quasars for which significant corrections for the Ly$\alpha$ forest and continuum absorption were made.   There appears to be a break in the continuum slope at around $1050{\rm \AA}$Ê with $\alpha_{\lambda}\approx-0.2$ in the far UV.  This composite spectra was used as one possible UV extension to the sub--912{\rm \AA} SEDs of the `SDSS', `RR1' and `RR2' AGN templates and can be seen in Fig. \ref{fig:agn} as the line labelled `UVHST'.
\item the `UVHST' is approximately a flat continuum.  In order to explore the two alternatives, the sub--912{\rm \AA} SEDs of the `SDSS', `RR1' and `RR2' AGN templates were also modelled as: a `UVrise' with $\alpha_{\lambda}$=$-$1.5; a `UVdrop' with $\alpha_{\lambda}$=+3.  These can also be seen in Fig. \ref{fig:agn}.  Recall that IGM absorption is applied to these SEDs depending on redshift.
\end{enumerate}
%%%%%%%%%%%%%%%%%%%%%%%%%%%%%%%%%%%%%%%%%%%%%%%%%%%%%%%%%%%%%%%%%%%%%%%
%%%%%%%%%%%%%%%%%%%%%%%%%%%%%%%%%%%%%%%%%%%%%%%%%%%%%%%%%%%%%
\section{Spectroscopic comparisons}
\label{sec:specz}
An important stage in the development of any photometric redshift code is to run it on a catalogue of objects for which the redshifts are already known from spectrocopic observations.  A version of the ImpZ code has already used to study the evolution of the UV radiation density, the dust opacity, and hence the SFH for galaxies in the HDF North and South (RR03).  As part of this, a comparison was made between the photometric output and the spectroscopic redshifts of 152 HDF North galaxies.  For galaxies with at least 4 photometric bands (U,B,V,I) it was found that the spectroscopic redshifts were successfully matched to an accuracy of around 10 per cent in (1+z).  Around 2.5 per cent (defined where there was one or more secondary minima with $\chi^2$ less than 1.0 above the global minimum and (1+z) differing by more than 20 per cent) of galaxies were found to have a `significant' redshift alias .

The version of ImpZ in RR03 has also been used to generate photometric redshifts for the final band-merged ELAIS catalogue (6.7, 15, 90 and 175$\mu$m, and $U$, $g\arcmin$, $r\arcmin$, $i\arcmin$, $Z$, $J$, $H$, $K$ and 20cm; \citealt{RR2004MNRAS.345..1290R}), fitting both galaxies and quasars by using their $U$, $g\arcmin$, $r\arcmin$, $i\arcmin$, $Z$, $J$, $H$, and $K$ data.  Again the photometric redshifts were accurate to around 10 per cent in (1+z), with a greater dispersion in $\Delta z$ for AGN fits.

In order to extend the applicability of ImpZ, this investigation explores Impz's application to two further sets of photometric data with spectroscopic redshifts.
%%%%%%%%%%
\subsection{ELAIS spectroscopic redshifts}
\label{subsec:wfsspeccat}
The spectroscopic data in the ELAIS N1 and N2 fields (P\'{e}rez--Fournon et al. 2003; in prep.) is comprised of two samples: 

a) objects observed with the William Herschel Telescope and the fibre-fed WYFFOS/Autofib2 spectrograph at the Observatorio del  Roque de Los Muchachos, La Palma, as part of an International Time Project approved by the Comit\'e Cient\'\i fico Internacional of the Observatories of the Instituto de Astrof\'\i sica de Canarias for the year 2000 (PI I. P\'erez-Fournon), and 

b) objects from the SDSS First Data Release (\citealt{Abazajian2003AJ....126.2081A}).

These data comprise 172 extragalactic sources, with spectroscopic redshifts ranging from a 0.0264 to 2.9426, and detections in 0 to 5 of the INT WAS bands.  The mean limiting magnitudes (Vega, 5$\sigma$ in 600 seconds) in each filter across these two areas are 23.40(U), 24.94(g$\arcmin$), 24.04(r$\arcmin$), 23.18(i$\arcmin$), and 21.90(Z).  The spectroscopic redshift information also includes a `QSO'/`GALAXY' flag (26/146 objects).  These data are reduced to 163 by removing objects flagged as saturated via the INT WAS class flags (1 source), objects that have no WAS detections (4 sources), objects with only one detection (3 sources), and objects flagged as being contaminated by a bright star (1 source), leaving 157 objects with detections in 5 bands, and 6 with detections in 4 bands, sub-divided into 25 `QSO' and 138 `GALAXY' sources.  The `QSO'/`GALAXY' information for INT WAS data enables us to decide on the best combination of the INT WAS class flags for defining possible AGN's that should be included in AGN template-fitting. and additionally gives a direct measure of how well the galaxy and AGN templates manage to separate out these two populations.  Hence a choice can then be made of which AGN templates to use, and what treatment in the UV is most successful.
Sources are defined as $stellar$ (and so AGN templates are considered in addition to the galaxy templates) based on class flags.  The INT WAS sources are classed in each band as either $-$1: point--like, 0: noise, 1: non--stellar, $-$9: saturated, $-$2: could be stellar,  $-$3: might just be stellar.  The best bands to define stellarity are $g\arcmin$, $r\arcmin$ and $i\arcmin$ so if the flag is $-$1 in any of these bands then the source is defined as $stellar$.  The choice of this definition of $stellar$ was reached after extensive tests with the spectroscopic INT WAS ELAIS catalogue which found that this identified the most number of actual quasars whilst keeping galaxy contamination to a minimum.

This $stellar$ definition splits the sample into 24 $stellar$ and 139 $non$-$stellar$ sources.

Galactic extinction in ELAIS N1 and N2 is low (they were after all selected for their low 100$\micron$ intensity).  Both have extinctions of $E(B-V)=0.007$, which using the extinction--wavelength relation in \cite{Cardelli1989ApJ...345..245C} equates to A$_{\rm v}\approx0.022$ for R$_v$=3.1.  We can also use the same relation to find the extinction in each of the WAS bandpasses and correct for it (see Table \ref{table:galactic}).  The effect of varying R$_v$ on the shape of the extinction curve is most apparent at the shorter wavelengths, and along with uncertainties in the actual form of the average R$_v$--dependent extinction law, the accuracy of these corrections is around 0.002 in magnitude (less than errors in the actual photometry).
%%%%%%%%%%
\subsection{CDFN spectroscopic redshifts}
\label{subsec:cdfnspeccat}
The \cite{Barger2002AJ....124.1839B} catalogue in the CDFN was also used.  This comprises an X--ray selected catalogue (from the 1Ms Chandra observation of the HDF North; \citealt{Brandt2001AJ....122.2810B}) of 169 objects with spectroscopic redshifts and broad band photometry ($B$, $V$, $R$, $I$, $z\arcmin$ from Subaru/Suprime Cam and a notched $HK\arcmin$ filter with a central wavelength of 1.8$\mu$m from the University of Hawaii 2.2m telescope/QUIRC -- see Fig. \ref{fig:igmfilters} for filter transmissions combined with the CCD responses).  The resolution of the X--ray and optical/near-IR observations are similar (around 1$\arcsec$) with nearly all true counterparts expected to lie within the 2$\arcsec$ (5$\sigma$) radii of the X--ray source for sources within 6.5$\arcmin$ of the approximate X--ray image centre and 3.6$\arcsec$ radii (4$\sigma$) for sources beyond this radius.  The 1$\sigma$ limiting magnitudes are (AB magnitudes): $B$(29.0), $V$(28.5), $R$(29.2), $I$(27.6), $z\arcmin$(27.0) and $HK\arcmin$(23.3, Vega).  As no errors are provided with these measurements, photometric errors of 0.05 in the Subaru bands and 0.15 in $HK\arcmin$ have been assumed.  The catalogue is reduced to 161 sources after removal of those flagged as saturated (7 sources) or contaminated (1 source).  In addition, $HK\arcmin$ detections were only used if they were brighter than magnitude 20.0 for data quality reasons.  This means there are 105 sources with detections in 6 bands, 52 with 5 bands and 4 with four bands.  Dropout treatment was applied to the two shortest wavebands -- B and V.

Since this sample is X--ray selected, we can expect a large proportion to be AGN, or AGN--dominated.  Hence although this sample does not have an exactly analogous set of filters to the INT WAS, it is another excellent test-bed for the AGN template-fitting.  It also has a larger sample of high-redshift galaxies which test the ImpZ application to galaxies across a broader redshift range.  In place of the $stellar$ definition in the application to the INT WAS survey, AGN template-fitting was carried out in addition to the usual galaxy template-fitting for objects that are flagged as being optically compact [C -- 23 sources], having broad-line features [B -- 6 sources], or having both [BC -- 23 sources].  Hence 52 of 161 sources have AGN template-fitting applied.  \cite{Barger2002AJ....124.1839B} used HYPERZ on this sample but only managed to get about 1/3 of their broad-line sources' photometric redshifts within 25 per cent of the spectroscopic values.

Galactic extinction in CDFN is again low, with extinction $E(B-V)$=0.012, which using the extinction--wavelength relation in \cite{Cardelli1989ApJ...345..245C} equates to A$_{\rm v}\approx0.037$ for R$_v$=3.1.  We can also use the same relation to find the extinction in each of the bandpasses and correct for it (see Table \ref{table:galactic}).%  The uncertainty in the Galactic extinction corrections for the CDFN sample is around 0.003, larger than for the INT WAS ELAIS sample due to the larger $E(B-V)$ extinction along the $l.o.s.$ to the CDFN region. 
%%%%%%%%%%%%%
\begin{table}
\caption{\scriptsize{ELAIS N1 and N2 Galactic extinction corrections for the INT WAS filters and CDFN Galactic extinction corrections for the Barger et al. (2002) filter set, derived using Cardelli et al. (1989) with R$_v$=3.1.% and are accurate to around 0.002 for ELAIS N1 and N2 and around 0.003 for CDFN, allowing for uncertainty in R$_v$.
}}
\begin{tabular}{|c c c c c c c|}
\hline
\hline
$INT$ &$WAS$ &$filters$ &&&&\\
\hline
 $filter$& $U$ & $g\arcmin$ & $r\arcmin$ & $i\arcmin$ & $Z$&\\
  \hline
A$_{\lambda}$ & 0.034 & 0.025 & 0.019 & 0.013 & 0.010&\\
\hline
\hline
$Barger$& $et$ $al.$ &$filters$&&&&\\
\hline
$filter$ & $B$ & $V$ & $R$ & $I$ & $z\arcmin$ & $HK\arcmin$\\
\hline
A$_{\lambda}$ & 0.049 & 0.037 & 0.031 & 0.022 & 0.018 & 0.005\\
\hline
\hline
\end{tabular}
\label{table:galactic}
\end{table}
%%%%%%%%%%
\subsection{Comparison to HYPERZ}
\label{subsec:hyperzcomp}
In order to see how the popular template-fitting photometric code HYPERZ compares, it has been run on the same two catalogues, using as similar parameters as possible.  The set of 6 galaxy templates was used, with RR1UVrise and RR2UVrise used for AGN-fitting.  %Since HYPERZ does not except a $stellar$ flag, the samples were split in two ($stellar$ and $non$--$stellar$ sources) and HYPERZ run separately on each, using all 8 SEDs for the $stellar$ samples and only the 6 galaxy SEDs for the $non$--$stellar$ samples.  Results were then re-combined for analysis.
  One difference between the HYPERZ and ImpZ codes is that the A$_v$ prior that increases the $\chi^{2}_{red}$ for increasing $|$A$_v|$ is not implementable in HYPERZ.  Similarly the prior that makes AGN fits more preferable than galaxy fits for $stellar$ sources is not available in HYPERZ (these priors were set out in \S\ref{subsec:code}).  HYPERZ was run with the same redshift range, but a redshift step in (1+z), rather than in log$_{10}$(1+z).  The same absolute magnitude limits as in ImpZ were used (but the equivalent limits in $g\arcmin$ band for INT WAS) for the galaxy and AGN fits, with an A$_v$ range of 0.0 to 1.0 for galaxy fits and A$_v$=0 for AGN and elliptical fits.  Two reddening laws were tried -- \cite{Calzetti2000ApJ...533..682C} and \cite{Seaton1979MNRAS.187P..73S} (fit by \citealt{Fitzpatrick1989IAUS..135...37F} for Milky Way), with galaxies getting the best results with the \cite{Seaton1979MNRAS.187P..73S} law.  Allowing A$_v$ to vary for AGN fits gave very similar results to setting A$_v$=0, with the best results shown in Fig. \ref{fig:specz} and Fig. \ref{fig:specz2}.  The results of this comparison to HYPERZ are in \S\ref{subsec:hyperzresults}.
%%%%%%%%%%%%%%%%%%%%%%%%%%%%%%%%%%
\section{Spectroscopic redshift results}
\label{sec:specresults}
We measure the reliability and accuracy of the photometric redshifts via the fractional error $\Delta z/(1+z)$ for each source, examining the mean error $\overline{\Delta} z/(1+z)$, the $rms$ scatter $\sigma_z$ and the rate of `catastrophic' outliers $\eta$, defined as the fraction of the full sample that has $|\Delta z/(1+z)|>0.2$.

$\Delta z/(1+z)$, $\overline{\Delta} z/(1+z)$ and $\sigma_{z}$ are calculated as follows:

\begin{equation}
\label{eqn:del_z}
\Delta  z/(1+z)= \left(\frac{z_{phot}-z_{spec}}{1+z_{spec}}\right)
\end{equation}
and
\begin{equation}
\label{eqn:del_zmean}
\overline{\Delta} z/(1+z)= \sum\left(\frac{z_{phot}-z_{spec}}{1+z_{spec}}\right)/N.
\end{equation}
and
\begin{equation}
\label{eqn:rms}
\sigma_{\rm z}^2=  \sum\left(\frac{z_{phot}-z_{spec}}{1+z_{spec}}\right)^2/N.
\end{equation}
with N being the number of sources with both spectroscopic redshifts and photometric redshifts.

 %Conversion to a  per cent accuracy, P, in (1+z) is via  P=100(10$^{ \sigma_{\rm z}}$-1).\\
For both spectroscopic samples, the outlier--clipped $rms$, $\sigma_{red}$, calculated from sources with $|\Delta z/(1+z)|<0.2$ was around 0.07 and the outlier fraction was 4.9 per cent for the INT WAS ELAIS and 12.4 per cent for the CDFN sample.
\subsection{Results of ELAIS spectroscopic study}
\label{subsec:wfsresults}
\begin{figure*}
\begin{center}
\includegraphics[height=20cm,width=17cm]{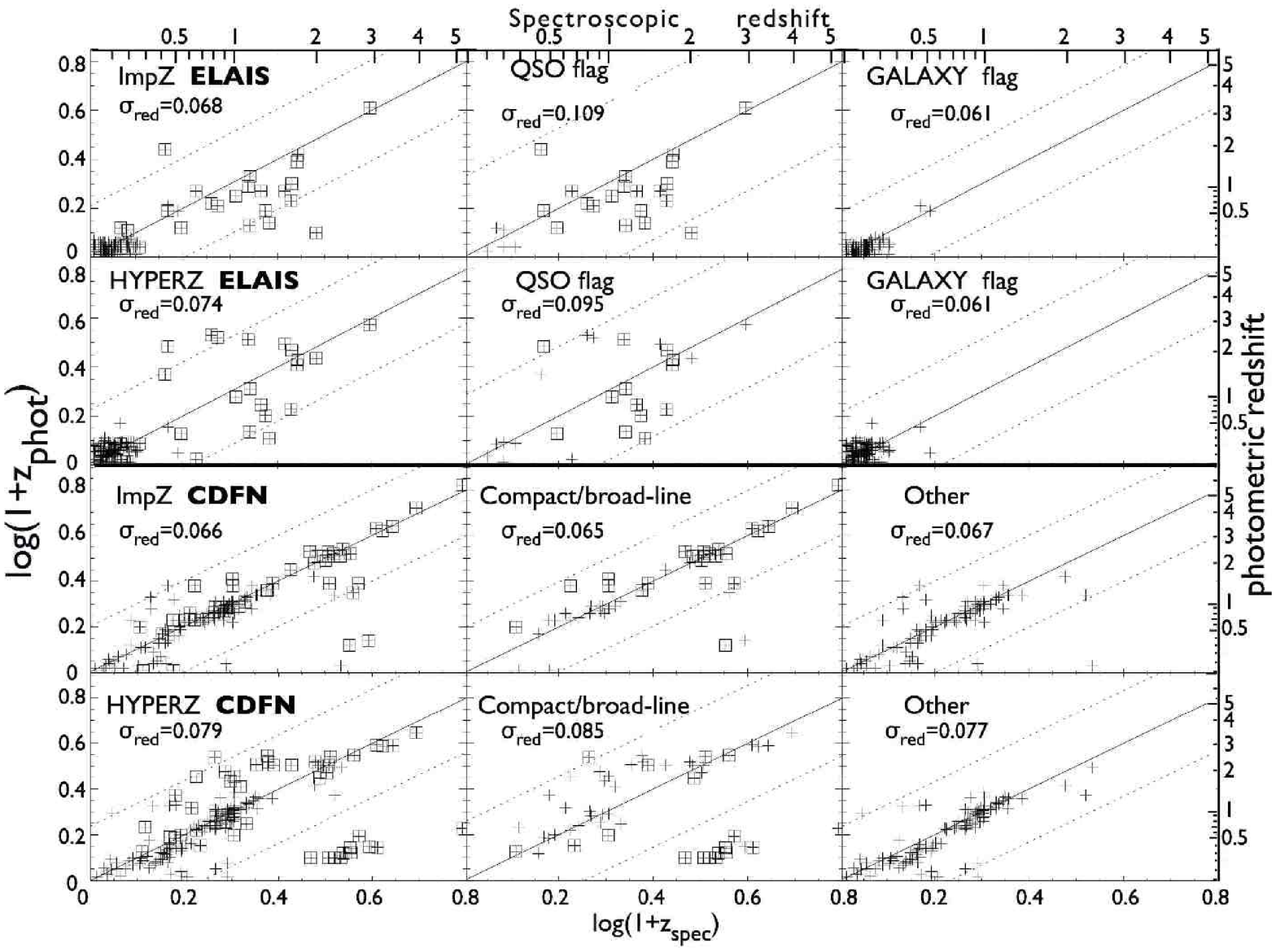}
\caption{\scriptsize{Spectroscopic validation results for the INT WAS ELAIS and CDFN spectroscopic samples.  The $rms$ values shown are the outlier-clipped $rms$ values, $\sigma_{red}$.  Top two rows shows INT WAS results (first row is ImpZ results, second row is HYPERZ results), set out as follows: first plot is all the data (crosses) with those flagged as `QSO' outlined with a square.  The dotted lines are the 3$\sigma_{red}$ limits; second plot is only `QSO' objects (crosses) with those fit as AGN outlined with a square.  Dotted lines are again 3$\sigma_{red}$; fouth plot are `GALAXY' objects (crosses) with those fit as AGN outlined by a square.  Dotted lines are 3$\sigma_{red}$ limits.  The third and fourth rows shows results from the CDFN sample with a similar format as to the first row: first plot is all the data (crosses) with those flagged as `B', `C' or `BC' outlined with a blue square.  The dotted lines are the 3$\sigma_{red}$ limits; third plot is only `B', `C' or `BC' objects (crosses) with those fit as AGN outlined with a square.  Dotted lines are again 3$\sigma_{red}$; fourth plot shows remaining objects (crosses) with dotted lines the 3$\sigma_{red}$ limits (third row is ImpZ results, fourth row is HYPERZ results).}}\label{fig:specz}
\end{center}
\end{figure*}

The code parameters and templates that gave the best results are described in \S\ref{subsec:photzsummary}.  Fig. \ref{fig:specz} shows the results of running ImpZ on the INT WAS ELAIS spectroscopic redshift sample with this `optimum' set of parameters and templates, with Fig. \ref{fig:specz_zoom} showing in close-up the behaviour at low redshift.  As well as plotting log$_{10}$(1+z$_{spec}$) versus log$_{10}$(1+z$_{phot}$) for the whole sample, the results for just those with a spectroscopic `QSO' flag are shown, as is the `GALAXY' sub-sample.  This better illustrates the varying success of the code on these different groups of sources, and the template-types that were best-fitting.  In particular, it can be seen that 20 of the 25 `QSO' objects were best-fitted by AGN templates. Four of the five that weren't were defined as $non$--$stellar$ following the definition set out in \S\ref{subsec:code} (so AGN-fitting was not applied to them) and all five are at low redshift ($z<0.3$).

ImpZ found solutions for all but one of the 162 sources.  The failed source is discussed in \S\ref{subsubsection:wfsoutliers}.

For the sample as a whole, the total $rms$ scatter, $\sigma_{tot}$, was 0.12, with $\overline{\Delta} z/(1+z)=-0.02$.  The outlier-clipped $rms$, $\sigma_{red}$ was 0.068 and the rate of `catastrophic' outliers, $\eta$ was 4.9 per cent.

The code was successful in identifying and fitting AGN templates to nearly all of the `QSO' objects, and was quite successful at returning an accurate  z$_{phot}$.  The total $rms$ scatter, $\sigma_{tot}$, was 0.29 for this sub-sample, with $\overline{\Delta} z/(1+z)=-0.11$.  The outlier-clipped $rms$, $\sigma_{red}$ was 0.11 and the rate of `catastrophic' outliers, $\eta$, was 32 per cent, higher than for the spectroscopic sample as a whole.  It can be seen from Fig. \ref{fig:specz} however that infact the agreement with z$_{spec}$ is reasonable (all but one are within the 3$\sigma_{red}$ limits), especially considering the infancy of applying photometric redshifts to quasars.   Nine of the 25 `QSO' sources were fit within 25 per cent of their z$_{spec}$ values.

The code was even better on the `GALAXY' sources, with $\sigma_{tot}$=0.061, $\overline{\Delta} z/(1+z)=-0.029$ and $\sigma_{red}$=0.061.  There were no `catastrophic' outliers.  Although these `GALAXY' sources were all relatively low redshift ($z<0.2$), the success is encouraging.

Statistical results for both the `best-case' setup and also for other ImpZ setups can be found in Table \ref{table:wfsresults}.
%%%%%%%%%%%%%%%%%%%%%%%%%%%%%%%%%%%%%%%%%%%%%%%%
%%%%%%%%%%%%%%%%%%%%%%%%%%%%%%%%%%%%%%%%%%%%%%%%
%%%%%%%%%%%%%%%%%%%%%%%%%%%%%%%%%%%%%%%%%%%%%%%%
\subsubsection{$Stellar$ definition}
\label{subsubsec:wfsstellar}
It is important to restrict which sources the AGN templates are fit to.  If the AGN templates are fit to all sources, instead of only fitting to those defined as $stellar$, then the overall accuracy of the ImpZ code on the INT WAS spectroscopic sample drops to $\sigma_{tot}$=0.16 and $\eta$ rises to 6.8 per cent.  This is because the inclusion of AGN templates introduces more degeneracies into the color-redshift space and we therefore wish to use further information, in this case the class flags, to break some of these degeneracies.

Similarly, although the $stellar$ sources are more likely to be quasars, if we restrict the fitting of galaxy templates to $non$--$stellar$ objects only, then the accuracy of ImpZ results for $stellar$ objects is reduced because not all of them will be quasars.  If galaxy templates are only fit to $non$--$stellar$ sources then $\sigma_{tot}$=0.16, $\overline{\Delta} z/(1+z)=0.01$, $\sigma_{red}$=0.07 and $\eta$=7.5 per cent.  Instead, galaxy templates are fit to $stellar$ objects, but AGN template--fits are made preferable through the prior set out in \S\ref{subsec:code}.  The alteration of this prior is discussed in \S\ref{subsubsec:relaxation}.

Bearing in mind that the code is to be applied to the INT WAS ELAIS survey as a whole where there will be proportionally less quasars and more galaxies, we wish to make the criteria for a $stellar$ definition and ensuing AGN-fitting as tight as possible in order to minimize galaxy contamination (and improve efficiency).  The best combination of class flags for identifying quasars without undue contamination from galaxies was found to be $stellar$ if the class flag is $-$1 in $g\arcmin$, $r\arcmin$, or $i\arcmin$ band.  This defines 21/25 `QSO' sources and 3/138 `GALAXY' sources as $stellar$.
%%%%%%%%%%%%%%%%%%%%%%%%%%%%%%%%%%%%%%%%%%%%%%%%
\subsubsection{AGN fitting}
\label{subsubsec:wfsagn}
The choice of which AGN templates to include in the template-set, and to a lesser extent how to treat them in the UV, had a large effect on the success of the code in obtaining accurate redshifts for $stellar$ objects, and also in whether the `QSO' sources were correctly identified as such and best-fitted by an AGN rather than a galaxy template.  At high enough redshifts for the different UV treatments to enter the INT WAS bandpasses ($z\ga2.5$) there are relatively few sources available to test the different behaviours -- the CDFN sample is more useful (see \S\ref{subsubsec:cdfnagn}).  For the highest redshift INT WAS source (`QSO' at $z=2.94$) it was found however that the AGN templates were not the best-fitting when the `UVHST' behaviour was used, with instead a low redshift Sbc template being fit.

The `UVdrop' behaviour was found to be reasonable, with the high-redshift source being fit as an AGN at $z=3.4$ whilst the `UVrise' gave the redshift as $z=3.1$.  The `UVrise' behaviour was therefore chosen to be the more successful, although the main reason for its choice was determined by the CDFN results (\S\ref{subsec:cdfnresults}) which had more high-redshift sources.

The `SDSS' AGN template was found to best-fit a reasonable number of $stellar$ sources but unfortunately the resulting photometric redshifts were inaccurate, with a larger number of catastrophic outliers.  For example, using the SDSS template with the `UVrise' behaviour best-fit 19/25 `QSO' sources, but with an increased $\sigma_{tot}$=0.30 and an outlier fraction of more than half.

%The Red AGN template only gave a good fit to one low-redshift source when it was included in the template-set.  This source was previously being fit as a low-redshift elliptical and the reason for the degeneracy between the two is that at low redshifts the INT WAS filter-set samples the same steep part of the two templates, making it hard to differentiate between the two (IR treatment does not have a bearing on application to the INT WAS ELAIS sample due the filters used and UV treatment will only have an effect at larger redshifts).%  The Red AGN will be still be included in the template--set for application to the INT WFS survey as a whole in the hope of finding high--redshift reddened AGN.

It was found that the best combination of AGN templates was to use the RR1 and RR2 templates with the UVrise behaviour.  Using just one or the other tended to increase the $rms$ scatter of the `QSO' objects from $\sigma_{tot}$=0.28 to roughly 0.40 and to increase the number of outliers.

It is noted that if no AGN templates are used $-$ only the 6 galaxy templates $-$ then the outlier fraction for the `QSO' sources increases to $\eta$=80 per cent with $\sigma_{tot}$=0.48.   Indeed, no source with $z_{spec}>$0.6 is fit within 3$\sigma_{tot}$ of z$_{spec}$.
%%%%%%%%%%%%%%%%%%%%%%%%%%%%%%%%%%%%%%%%%%%%%%%%
\subsubsection{High redshift sources}
\label{subsubsec:wfshighz}
Since the main power of photometric redshifts is in their application to higher redshifts, it is informative to examine the higher redshift sources as a separate sample.  Taking our redshift cut as z$_{spec}>0.3$ (sources below this redshift are plotted in close-up in Fig. \ref{fig:specz_zoom}) means that there are 22 sources in the high redshift sample, comprising 2 `GALAXY' and 20 `QSO' sources.  Hence the statistics for this sample are similar to that for the QSO sources alone, with $\sigma_{tot}$=0.34, $\overline{\Delta} z/(1+z)=0.12$ and $\sigma_{red}$=0.11 with an outlier rate of 36 per cent.  As would be expected, then, the accuracy for higher redshift sources is less than for lower redshift sources, but is still good.  A better grasp of high redshift performance is provided by the CDFN sample, which has more objects at z$_{spec}>0.3$ -- see \S\ref{subsubsec:cdfnhighz}.
%%%%%%%%%%%%%%%%%%%%%%%%%%%%%%%%%%%%%%%%%%%%%%%
\subsubsection{Relaxation of parameters}
\label{subsubsec:relaxation}
\paragraph{A$_{v}$}%  Av=0 vs. Av free.
\label{para:wfsav}
If A$_{v}$ fitting is completely turned off, so that only A$_{v}$=0 solutions are allowed then the majority of the increase in $rms$ comes from increased scatter in the `GALAXY' sub-sample.    Hence the inclusion of A$_{v}$ as a parameter improves the accuracy of the redshifts, whilst presumably giving some information about the extinction of each source.  Several different A$_{v}$ limits were tried for the galaxy templates.  For example, since the ELAIS fields were originally IR--selected we might expect some sources to have high extinction so a range of A$_{v}=0$ -- 3 was used, with $\beta=3$ instead of 2 in the A$_{v}$ prior that minimises $\chi^{2}_{red}$ + $\beta$A$_v^2$.  This did not alter the results..  However, extending A$_{v}$ freedom to AGN templates increases the the outlier--rate for `QSO's to $\eta=40\%$.

\paragraph{AGN prior}
\label{para:wfsagnprior}
If the prior that a galaxy-fit to a $stellar$ source must have $\chi^{2}_{red}.^{galaxy}<(\chi^{2}_{red}.^{AGN}-4)$ is removed then only 17 of 25 `QSO' sources are fit as AGN and the highest redshift source (`QSO' at $z=2.94$) is instead fit as a low-redshift Sbc galaxy, with an overall increase in the $rms$ and outlier fraction.  If the prior is reduced in strength, to $\chi^{2}_{red}.^{galaxy}<(\chi^{2}_{red}.^{AGN}-2)$ then although the overall statistics are almost as good, the highest redshift source is again fit as a low-redshift Sbc galaxy.

\paragraph{Absolute Magnitude limits}
\label{para:wfsmag}
The choice of the absolute $B$-band magnitude, M$_B$, limits of the photometric solutions is important for removing physically unlikely low-redshift low-luminosity and high-redshift high-luminosity solutions which greatly increase the number of catastrophic outliers.  For example, allowing brighter AGN limits of --29$<$M$_B$$<$--17 instead of --27$<$M$_B$$<$--17 resulted in a number of high photometric redshift ($z>5$) fits to low redshift sources.  Similarly, increasing the galaxy limits to --25.5$<$M$_B$$<$--13 replaced a number of correct faint low-redshift fits with incorrect brighter high-redshift fits.  Incorrect low-redshift solutions for high-redshift souces were found to increase if the opposite action was taken and brighter galaxy and AGN fits were not allowed.  The use of an upper M$_B$ envelope that increased with redshift was found to be more successful than using a fixed value for galaxies, as might be expected from the known strong evolution in galaxy luminosities.  The selected redshift dependence used here was [$-22.5-2log_{10}(1+z)]<$M$_B<-13.5$, though [$-22.5-z/3]<$M$_B<-13.5$ worked equally well.\\
The final absolute magnitude limits were chosen in order to give the best agreement with the spectroscopic data whilst taking into consideration that the INT WAS ELAIS survey as a whole will have a greater diversity of sources.
%%%%%%%%%%%%%%%%%%%%%%%%%%%%%%%%%%%%%%%%%%%%%%%%%%%%%%%%%%%
\subsubsection{Outliers}
\label{subsubsection:wfsoutliers}
ImpZ fails to find a solution for one INT WAS ELAIS source, a `GALAXY' at redshift 0.1454 with a $non$--$stellar$ flag.  This source is not detected in $g\arcmin$, but is relatively bright in other bands (19.42, 18.05, 17.18, 16.75 in $U$, $r\arcmin$, $i\arcmin$ and $Z$).  It was flagged as having multiple counterparts so this might imply that the photometry is incorrect.  HYPERZ manages to find a solution at z$_{hyp}$=0.19 as an Sab with A$_{v}$=0.4.%  The success is likely to be the different ways in which the two codes treat drop--outs/non--detections.

The remaining most conspicious outliers are a group of three sources with high spectroscopic redshifts and lower photometric redshift solutions, falling below the 3$\sigma_{red}$ boundaries.  It is noted that two of these sources are at redshifts where the 4000{\rm \AA} break has left the longest INT WAS waveband (at $z\approx1.4$) but the 912{\rm \AA} Lyman limit has still not entered the shortest waveband ($z\approx2.5$).  Sources in this redshift range are harder to successfully derive redshifts for since the primary SED features on which photometric redshift-fitting relies are not available.  Table \ref{table:features} details the redshifts where common spectral features enter and leave the bandpasses used in the INT WAS and CDFN catalogues.  For five other sources that fall into this redshift range, ImpZ is still successful in deriving the correct redshift, implying a success-rate in this redshift region of 71 per cent.

The other outlier (a `QSO') lies above the 3$\sigma_{red}$ boundaries, with a low z$_{spec}$ and a high z$_{phot}$ solution as an AGN.  A solution near the correct redshift is found if the SDSS AGN template is used, however including this template degrades the performance of the code for the rest of the sample.
%%%%%%%%%%%%%%%%%%%%%%%%%%%%%%%%%%%%%%%%%%%%%%
\begin{figure}
\begin{center}
\includegraphics[height=6.5cm,width=8.5cm]{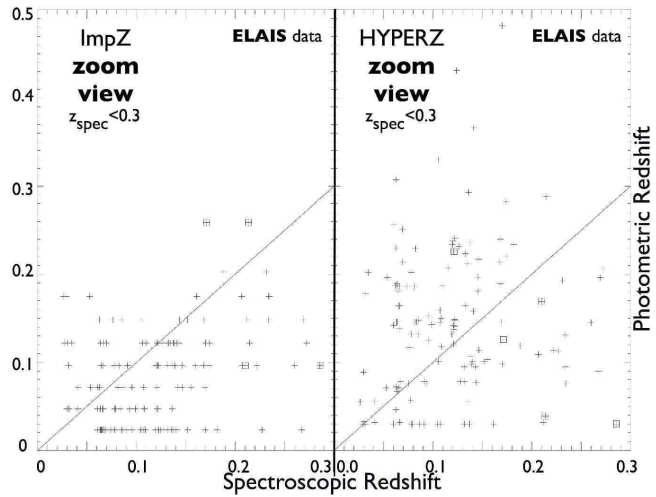}
\caption{\scriptsize{Zoomed view of spectroscopic validation results for the INT WAS ELAIS spectroscopic sample.  First plot shows all ImpZ data for $z_{spec}<0.3$ (crosses) with those flagged as `QSO' outlined with a square; second plot is the same plot for the HYPERZ results.  For both plots the scatter appears large since the photometric redshift method is accurate to perhaps 0.05 in (1+z) -- of order of (1+z) for low redshfits.}}\label{fig:specz_zoom}
\end{center}
\end{figure}
%%%%%%%%%%%%%%%%%%%%%%%%%%%%%%%%%%%%%%%%%%%%%%%%
%%%%%%%%%%%%%%%%%%%%%%%%%%%%%%%%%%%%%%%%%%%%%%%%
\begin{figure*}
\begin{center}
\includegraphics[height=16.cm,width=9.cm,angle=90]{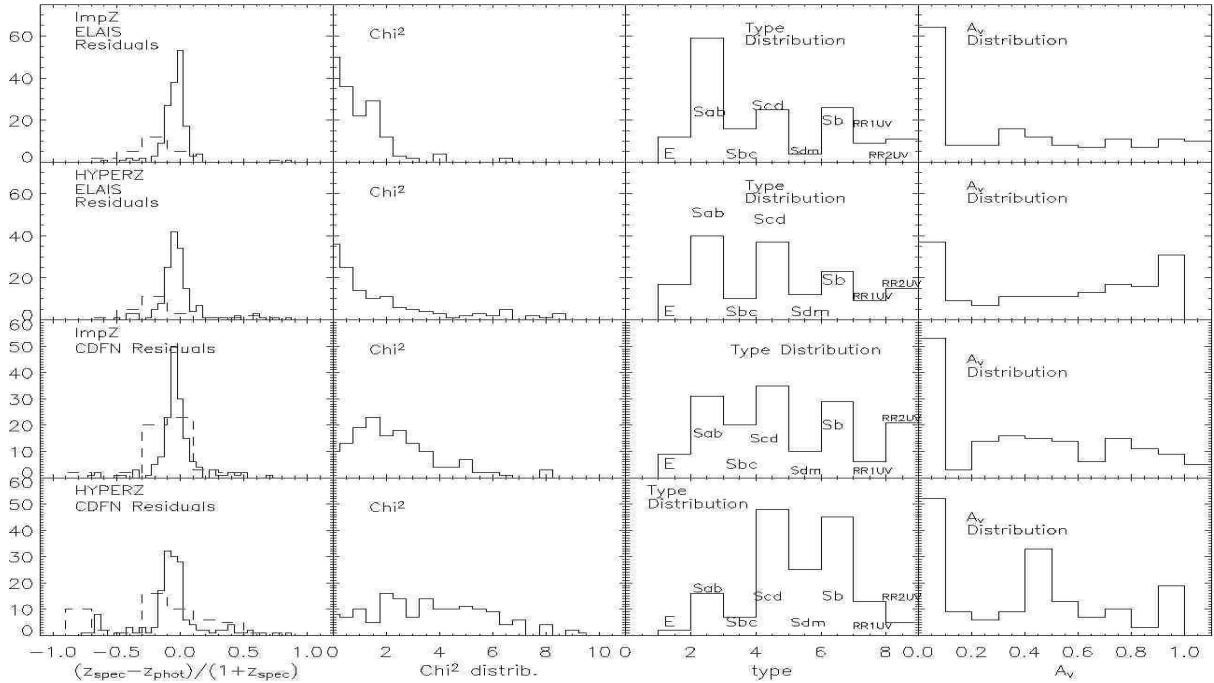}
\caption{\scriptsize{Spectroscopic validation results for the INT WAS ELAIS and CDFN spectroscopic samples.  Top two rows show INT WAS results (first row shows ImpZ results, second row shows HYPERZ results), set out as follows: first plot is the distribution of residuals between z$_{spec}$ and z$_{phot}$, normalised to (1+z$_{spec}$).  Solid histogram is for the whole sample, dashed histogram is that of QSO objects only; second plot shows the reduced chi$^2$ distribution of the fits; third plot is the type distribution of the fits where 1=E, 2=Sab, 3=Sbc, 4=Scd, 5=Sdm, 6=Sb, 7=RR1UVrise and 8=RR2UVrise; fourth plot is the A$_{v}$ distribution.  The third and fourth rows show results from the CDFN sample with the same format as the first two rows except that the dashed histogram is now that of broad-line and/or compact sources only (third row shows ImpZ results, fourth row shows HYPERZ results).}}\label{fig:specz2}
\end{center}
\end{figure*}
%%%%%%%%%%%%%%%%%%%%%%%
%TABLE OF RESULTS HERE
\begin{table*}%[!ht]
\caption{\scriptsize{Statistical results for application of the ImpZ code to the INT WAS ELAIS catalogue, for different ImpZ setups.}}
\centering
\scriptsize
\begin{tabular}{ c*{14}{c}}
\hline
\hline
\multicolumn{14}{c}{INT WAS ELAIS Statistics for different ImpZ setups}\\
\hline
\hline
\multicolumn{4}{c}{All sources} &\multicolumn{4}{c}{`QSO' sources} &\multicolumn{4}{c}{`GALAXY' sources}\\
 \multicolumn{1}{c}{$\overline{\Delta} z/(1+z)$}&\multicolumn{1}{c}{$\sigma_{tot}$}&\multicolumn{1}{c}{$\sigma_{red}$}&\multicolumn{1}{c}{$\eta$(\%)}
&&\multicolumn{1}{c}{$\overline{\Delta} z/(1+z)$}&\multicolumn{1}{c}{$\sigma_{tot}$}&\multicolumn{1}{c}{$\sigma_{red}$}&\multicolumn{1}{c}{$\eta$(\%)}
&&\multicolumn{1}{c}{$\overline{\Delta} z/(1+z)$}&\multicolumn{1}{c}{$\sigma_{tot}$}&\multicolumn{1}{c}{$\sigma_{red}$}&\multicolumn{1}{c}{$\eta$ (\%)}\\
%\cline{2-17}
\hline
\hline
\multicolumn{14}{c}{`Best--case' ImpZ setup as in \S5.1}\\
$-$0.02&0.12&0.068&4.9&&  $-$0.11&0.29&0.11&32.0& &$-$0.029&0.061&0.061&0.0\\
\hline
\multicolumn{14}{c}{AGN fit to all sources as in \S5.1.1}\\
$-$0.01&0.16&0.07&6.8& &$-$0.05&0.36&0.11&40.0&& $-$0.002&0.07&0.06&0.7\\
\hline
\multicolumn{14}{c}{Galaxies only fit to $non$--$stellar$ sources as in \S5.1.1}\\
0.01&0.16&0.07&7.5&& $-$0.08&0.32&0.11&36.0& &0.03&0.11&0.06&2.2\\
\hline
\multicolumn{14}{c}{SDSS template with UVrise as the only AGN template, as in \S5.1.2}\\
$-$0.04&0.13&0.07&8.0&& $-$0.21&0.30&0.11&52.0&& $-$0.004&0.06&0.06&0.0\\
\hline
\multicolumn{14}{c}{RR1 template with UVrise as the only AGN template, as in \S5.1.2}\\
$-$0.01&0.16&0.06&6.2&& $-$0.07&0.39&0.10&40.0& &$-$0.004&0.06&0.06&0.0\\
\hline
\multicolumn{14}{c}{RR2 template with UVrise as the only AGN template, as in \S5.1.2}\\
$-$0.03&0.15&0.07&8.0& &$-$0.17&0.36&0.12&52.0& &$-$0.004&0.06&0.06&0.0\\
\hline
\multicolumn{14}{c}{UV drop behaviour in place of UVrise, as in \S5.1.2}\\
$-$0.02&0.12&0.07&4.9&& $-$0.11&0.29&0.12&32.0&& $-$0.004&0.06&0.06&0.0\\
\hline
%\multicolumn{14}{c}{With the Red AGN as well as RR1 and RR2, as in \S5.1.2}\\
%$-$0.02&0.12&0.07&4.9&& $-$0.11&0.28&0.11&32.0& &$-$0.004&0.06&0.06&0.0\\
%\hline
\multicolumn{14}{c}{The 4 CWW templates in place of the E, Sab, Sbc, Scd, and Sdm templates, as in \S6.2}\\
$-$0.02&0.13&0.07&5.6&& $-$0.12&0.28&0.12&32.0&& $-$0.01&0.07&0.07&0.7\\
\hline
\multicolumn{14}{c}{Sab, R1UVrise and RR2UVrise as the only templates, as in \S6.2}\\
$-$0.05&0.14&0.07&8.6&& $-$0.19&0.30&0.12&48.0&& $-$0.03&0.07&0.06&1.5\\
\hline
\multicolumn{14}{c}{No AGN templates, as in \S5.1.2}\\
$-$0.07&0.20&0.06&12.3&& $-$0.42&0.48&0.11&80.0&& $-$0.004&0.06&0.06&0.0\\
\hline
\multicolumn{14}{c}{A$_{v}$ free for AGN templates}\\
$-$0.03&0.13&0.07&6.8&& $-$0.17&0.30&0.11&44.0& &$-$0.004&0.06&0.06&0.0\\
\hline
\multicolumn{14}{c}{A$_{v}$=0 to 3, with $\beta=3$ in A$_{v}$ prior (see \S\ref{subsec:code})}\\
$-$0.02&0.12&0.07&4.9&& $-$0.11&0.28&0.11&32.0&& $-$0.01&0.06&0.06&0.0\\
\hline
\multicolumn{14}{c}{A$_{v}=-0.4$ to 1.0}\\
$-$0.02&0.13&0.07&6.2&& $-$0.11&0.29&0.12&36.0& &$-$0.004&0.06&0.06&0.7\\
\hline
\multicolumn{14}{c}{A$_{v}$ set to zero for all templates, as in \S5.1.3}\\
0.01&0.13&0.08&5.6&& $-$0.11&0.29&0.10&32.0&& 0.02&0.08&0.07&0.7\\
\hline
\multicolumn{14}{c}{AGN prior removed, as in \S5.1.3}\\
$-$0.01&0.15&0.07&7.4&& $-$0.17&0.34&0.12&44.0&& 0.02&0.07&0.06&0.7\\
\hline
\multicolumn{14}{c}{AGN prior set to 2 instead of 4, as in \S5.1.3}\\
$-$0.01&0.14&0.07&6.2&& $-$0.14&0.32&0.11&36.0& &0.02&0.07&0.06&0.7\\
\hline
\multicolumn{14}{c}{AGN prior set to 25 instead of 4, as in \S5.1.3}\\
0.003&0.14&0.07&6.2&& $-$0.08&0.32&0.11&36.0&& 0.02&0.07&0.06&0.7\\
\hline
\multicolumn{14}{c}{Without U band information, as in \S6.2}\\
0.01&0.43&0.06&13.0&& $-$0.11&0.43&0.11&60.0& &0.03&0.43&0.06&4.4\\
\hline
\multicolumn{14}{c}{Without Z band information, as in \S6.2}\\
$-$0.03&0.13&0.07&7.4& &$-$0.13&0.30&0.10&48.0&& $-$0.01&0.06&0.06&0.0\\
\hline
\multicolumn{14}{c}{Brighter AGN limits of --29$<$M$_B$$<$--17.5, as in \S6.3}\\
0.01&0.25&0.07&6.2& &0.10&0.64&0.12&40.0&& $-$0.004&0.06&0.06&0.0\\
\hline
\multicolumn{14}{c}{Reducing faint AGN limits to --27.0$<$M$_B$$<$--15, as in \S5.1.3}\\
$-$0.02&0.12&0.07&4.9& &$-$0.11&0.28&0.11&32.0&& $-$0.01&0.06&0.06&0.0\\
\hline
\multicolumn{14}{c}{Brighter galaxy limits to --23.5$<$M$_B$$<$--13.5, as in \S5.1.3}\\
$-$0.02&0.13&0.07&6.2&& $-$0.11&0.28&0.11&32.0& &$-$0.003&0.06&0.06&1.5\\
\hline
\multicolumn{14}{c}{Brighter galaxy limits to --25.5$<$M$_B$$<$--13.5, as in \S5.1.3}\\
0.01&0.33&0.06&7.4&& $-$0.10&0.29&0.12&36.0&& 0.03&0.34&0.05&2.2\\
\hline
\multicolumn{14}{c}{Reducing faint galaxy limits to --23$<$M$_B$$<$--17.5, as in \S5.1.3}\\
$-$0.02&0.12&0.06&4.9&& $-$0.11&0.28&0.11&32.0&& 0.001&0.06&0.06&0.0\\
\hline
\hline
\end{tabular}
\footnotesize
\label{table:wfsresults}
\normalsize
\end{table*}
%%%%%%%%%%%%%%%%%%%%%%%%%%%%%%%%%%%%%%%%%%%%%%%%
\subsection{Results of CDFN spectroscopic study}
\label{subsec:cdfnresults}
Fig. \ref{fig:specz} shows the results of running ImpZ on the CDFN spectroscopic redshift sample with the same `optimum' set of parameters and templates as plotted for the INT WAS sample.  As expected, since there is less available information on which to base the $stellar$ and $non$-$stellar$ definition and no actual `QSO' or `GALAXY' flag, the resulting analysis can be less clearly separated into results for galaxies and quasars.

ImpZ found solutions for all 161 sources.  The total $rms$ scatter, $\sigma_{tot}$, was 0.17, with $\overline{\Delta} z/(1+z)=-0.01$.  The outlier--clipped $rms$, $\sigma_{red}$ was 0.07 and the rate of `catastrophic' outliers, $\eta$ was 12.4 per cent.  From Fig. \ref{fig:specz} it can be seen that the code really only went badly wrong for 3 sources.

The code was successful in identifying and fitting AGN templates to 27/52 of the broad-line or compact objects, and overall was successful at returning an accurate  photometric redshift.  The total $rms$ scatter, $\sigma_{tot}$, was 0.19 for this sub-sample, with $\overline{\Delta} z/(1+z)=-0.01$.  The outlier-clipped $rms$, $\sigma_{red}$ was 0.07 and the rate of `catastrophic' outliers, $\eta$, was 17.3 per cent.

The code is also successful for the remaining sources (not B and/or C sources), with $\sigma_{tot}$=0.16, $\overline{\Delta} z/(1+z)=-0.01$ and $\sigma_{red}$=0.07.  The rate of `catastrophic' outliers, $\eta$, was 10.1 per cent.

Statistical results for both the `best-case' setup and also for other ImpZ setups can be found in Table \ref{table:cdfnresults}.
%%%%%%%%%%
\subsubsection{AGN fitting}
\label{subsubsec:cdfnagn}
Nineteen of 29 (66 per cent) broad-line (B or BC) source redshifts were within 25 per cent of their spectroscopic values.  As a comparison, \cite{Barger2002AJ....124.1839B} got 1/3 of the sample within this tolerance.

If we use just BC (broad-line and compact) or C (compact) as the $stellar$ definition (in place of B, C, BC) then the broad-line source numbered 174 in \cite{Barger2002AJ....124.1839B} (with the highest spectroscopic redshift, $z=5.19$) is incorrectly fit as a low-redshift galaxy.  The reason why the B flag should perhaps not be used in the $stellar$ definition is that this is a spectroscopically rather than photometrically derived property, and so would not be available to a photometric survey in general.

As with the INT WAS ELAIS spectroscopic sample, the most successful AGN templates were found to be the RR1 and RR2 templates and due to the greater number of high-redshift sources in the CDFN sample, the best UV treatment could investigated more clearly.  It was found that using the UVdrop or UVHST behaviours was less successful than UVrise.  The UVHST behaviour increased the population of low-redshift sources incorrectly fit as high-redshift, although the overall success was good, with the main result being a slight increase in the outlier fraction for the $stellar$ sources.  The UVdrop treatment had a similar effect, making the UVrise behaviour the most effective.

It is noted that if no AGN templates are included in the template-set then the outlier-rate for $stellar$ sources increases to 58 per cent, with a large population of high-redshift $stellar$ sources incorrectly placed at lower redshifts.  If AGN templates are fit along with galaxy templates to $non$-$stellar$ sources then the outlier-rate for $non$-$stellar$ sources increases to 28 per cent and $\sigma_{tot}$ rises to 0.88. If, conversely, galaxy templates are only fit to $non$-$stellar$ sources then the $rms$ scatter of the $stellar$ sources becomes much larger, with $\sigma_{tot}$=1.42.
%%%%%%%%%%%%%%%%%%%%%%%%%%%%%%%%%%%%%%%%%%%%%%%%
\subsubsection{High redshift sources}
\label{subsubsec:cdfnhighz}
Again, we examine the higher redshift sources as a separate sample.  Taking our redshift cut as z$_{spec}>0.3$ (sources below this redshift are plotted in close-up in Fig. \ref{fig:specz_zoom}) means that there are 143 sources in the high redshift sample, comprising 92 $non$-$stellar$ and 51 $stellar$ sources.  The statistics for this sample are $\sigma_{tot}$=0.18, $\overline{\Delta} z/(1+z)=0.01$ and $\sigma_{red}$=0.07 with an outlier rate of 13 per cent.  The accuracy for higher redshift sources is therefore still good.  The performance is better than for the high--redshift sample in the INT WAS ELAIS catalogue since in that case 90 per cent of the sources were `QSO', for which photometric redshift--fitting is less accurate.
%%%%%%%%%%%%%%%%%%%%%%%%%%%%%%%%%%%%%%%%%%%%%%%
%%%%%%%%%%%%%%%%%%%%%%%%%%%%%%%%%%%%%%%%%%
\subsubsection{Relaxation of parameters}
\label{subsubsec:cdfnrelaxation}
\paragraph{A$_{v}$}%  Av=0 vs. Av free.
\label{para:cdfnav}
If A$_{v}$ fitting is completely turned off, so that only A$_{v}$=0 solutions are allowed then $\eta$ increases to 17.4 per cent.  Again, the inclusion of A$_{v}$ as a parameter improves the accuracy of the redshifts, though analysis of a sample with known A$_{v}$ would be required to quantify how the resulting A$_{v}$ of the solution compares to actual A$_{v}$.

\paragraph{AGN prior}
\label{para:cdfnavagn}
If the prior that a successful galaxy-fit to a $stellar$ source must have $\chi^{2}_{red}.^{galaxy}<(\chi^{2}_{red}.^{AGN}-4)$ is removed then only 13 of the 52 $stellar$ sources are fit by AGN and many of those that are then fit as galaxies are placed at much lower z$_{phot}$.  Using a reduced prior of $\chi^{2}_{red}.^{galaxy}<(\chi^{2}_{red}.^{AGN}-2)$ causes the highest z$_{spec}$ source to again be fit as a low redshift galaxy, and increasing the prior to $\chi^{2}_{red}.^{galaxy}<(\chi^{2}_{red}.^{AGN}-25)$ creates a number of false high--z$_{phot}$ AGN solutions.

\paragraph{Absolute Magnitude limits}
\label{para:cdfnmag}
As with the INT WAS sample, allowing brighter AGN limits of --29$<$M$_B$$<$--17.5 instead of --27$<$M$_B$$<$--17.5 resulted in a number of high photometric redshift ($z>5$) fits to low redshift sources.  Similarly, increasing the luminous galaxy limits to  --25.5$<$M$_B$$<$--13.5 replaced a number of correct faint low-redshift fits with incorrect luminous high-redshift fits.  Incorrect low-redshift solutions for high-redshift souces were found to increase if the opposite action was taken and brighter galaxy and AGN fits were not allowed.  Again, allowing an increase in the upper galaxy M$_B$ envelope with redshift was found to be the best solution (i.e $-22.5-2log_{10}(1+z)$).
%%%%%%%%%%%%%%%%%%%%%%%%%%%%%%%%%%%%%%%%%%
\subsubsection{Outliers}
\label{subsubsection:cdfnoutliers}
There are a group of three high spectroscopic redshift sources whose photometric solutions are significantly lower than their spectroscopic redshifts (there are a total of six sources below the 3$\sigma_{red}$ boundary, and one above, but the others are close to the boundaries).  It is noted that this group of three are at redshifts where the 4000{\rm \AA} break has left the Z waveband (at $z\approx1.5$) but the 912{\rm \AA} Lyman limit has still not entered the shortest waveband ($z\approx3.3$).  Sources in this redshift are harder to successfully derive redshifts for since the primary SED features on which photometric redshift-fitting relies are not available.  For 20 other sources that fall into this redshift range, ImpZ is still successful in deriving the correct redshift, implying a success-rate in this redshift region of 87 per cent.  The slightly better performance than for the INT WAS ELAIS sample (71 per cent) for the corresponding `feature desert' is likely to be due to the inclusion of HK' information (which 4000{\rm \AA} enters at $z\approx2.7$) and partly due to the poor-number statistics for the INT WAS ELAIS sample (only 6 sources in that redshift range).
%%%%%%%%%%%%%%%%%%%%%%%%%%%%%%%%%%%%%%%%%%
\subsection{Statistical properties}
\label{subsec:wfsstats}
The distribution of residuals of the ImpZ solutions in Fig. \ref{fig:specz2} is quite strongly peaked around zero, with a Gaussian-like distribution and a slightly non-Gaussian extension to larger values.  The distributions show that the ImpZ code is successful to an accuracy of perhaps 0.1 to 0.2 in (1+z) for nearly all sources.  The distributions of ImpZ solutions (dashed line in \ref{fig:specz2}) for the `QSO' sub-sample is also Gaussian--like, but these distributions are wider and shallower.  For the INT WAS ELAIS `QSO's the distribution is peaked at around $-0.2$ rather than zero, implying that the redshift solutions tend to be slightly larger than the true values.  This could be because A$_{v}$ is set to zero for AGN so a QSO reddened by some extinction is fit as unreddened, but at a slightly higher redshift.  However, this is not seen in the larger CDFN sample.

The $\chi^{2}_{red}$ distributions of the ImpZ solutions shows that the majority of solutions are good, with $\chi^{2}_{red}$ peaking below 2.  There is no large population of poor--$\chi^{2}_{red}$ solutions.  From the ImpZ type distributions it can be see that Sab solutions dominate, with a sizeable fraction of Scd and starbursts and very few Sdm solutions.

The A$_{v}$ distributions are strongly peaked at zero with an even spread of higher values (but no clustering at the highest value of A$_{v}=1$ which would imply that the A$_{v}$ range was insufficient).
%%%%%%%%%%%%%%%%%%%%%%%%%%%%%%%%%%
\subsection{HYPERZ results}
\label{subsec:hyperzresults}
From Fig. \ref{fig:specz} and Fig. \ref{fig:specz2} it can be seen that for both the INT WAS and CDFN spectroscopic samples, the ImpZ code manages to do better.  The strongest cause of these differences are probably the AGN and A$_v$ priors that are used in the ImpZ code.  In particular, HYPERZ is not as successful for the broad-line and/or compact sources in the CDFN sample, with a larger population of high-redshift sources incorrectly placed at low-redshift (see Fig. \ref{fig:specz}).

Overall, HYPERZ has, for the INT WAS sample, $\sigma_{tot}$=0.17, $\overline{\Delta} z/(1+z)=0.02$, $\sigma_{red}$=0.073 and $\eta$=8.0 per cent.  For the CDFN sample the statistics were $\sigma_{tot}$=0.26, $\overline{\Delta} z/(1+z)=-0.04$, $\sigma_{red}$=0.08 and $\eta$=24.2 per cent.  Only 9/29 broad--line source redshifts were within 25 per cent of their spectroscopic values.

From Fig. \ref{fig:specz2} it can be seen that the non-Gaussian tails of the redshift residuals are more pronounced, particularly for the CDFN sample, and the $\chi^{2}_{red}$ of the solutions are only weakly peaked to low values.

A consideration of the `feature desert' in redshift space where the 4000{\rm \AA} break has left the optical wavebands, and the 912{\rm \AA} Lyman limit has not yet entered shows a marked difference from the ImpZ results for the CDFN, indeed it is this region where HYPERZ most strongly diverges from ImpZ in its outlier-rate:  for the CDFN sample, ImpZ had a 87 per cent success-rate with the 23 sources in this range, whereas HYPERZ fails for 10 (57 per cent success).  
%It's not as good...Possible reasons for this.  Any individual objects where it is much better.  Av + type distribs
%%%%%%%%%%%%%%%%%%%%%%%%%%%%%%%%%%%%%%%%%
\begin{table*}%[!ht]
\caption{\scriptsize{Statistical results for application of the ImpZ code to the CDFN catalogue, for different ImpZ setups.}}
\centering
\scriptsize
\begin{tabular}{ c*{14}{c}}
\hline
\hline
\multicolumn{14}{c}{CDFN Statistics for different ImpZ setups}\\
\hline
\hline
\multicolumn{4}{c}{All sources} &  &\multicolumn{4}{c}{Broad--line and/or Compact sources} &  &\multicolumn{4}{c}{Other sources}\\
\multicolumn{1}{c}{$\overline{\Delta} z/(1+z)$}&\multicolumn{1}{c}{$\sigma_{tot}$}&\multicolumn{1}{c}{$\sigma_{red}$}&\multicolumn{1}{c}{$\eta$(\%)}
&  &\multicolumn{1}{c}{$\overline{\Delta} z/(1+z)$}&\multicolumn{1}{c}{$\sigma_{tot}$}&\multicolumn{1}{c}{$\sigma_{red}$}&\multicolumn{1}{c}{$\eta$(\%)}
&  &\multicolumn{1}{c}{$\overline{\Delta} z/(1+z)$}&\multicolumn{1}{c}{$\sigma_{tot}$}&\multicolumn{1}{c}{$\sigma_{red}$}&\multicolumn{1}{c}{$\eta$ (\%)}\\
%\cline{2-17}
\hline
\hline
\multicolumn{14}{c}{`Best--case' ImpZ setup as in \S5.2}\\
$-$0.01&0.17&0.07&12.4&  &  0.01&0.19&0.07&17.3&  &  $-$0.01&0.16&0.07&10.1\\
\hline
\multicolumn{14}{c}{AGN fit to all sources as in \S5.2.1}\\
0.17&0.70&0.07&18.6& &  0.01&0.19&0.07&17.3& &  0.24&0.84&0.07&17.3\\
\hline
\multicolumn{14}{c}{Galaxies only fit to $non$--$stellar$ sources as in \S5.2.1}\\
0.25&0.80&0.07&24.8& &  0.86&1.42&0.07&56.3& &  $-$0.01&0.16&0.07&11.0\\
\hline
\multicolumn{14}{c}{SDSS template with UVrise as the only AGN template, as in \S5.2.1}\\
$-$0.07&0.23&0.07&22.4& &  $-$0.19&0.34&0.08&48.1& &  $-$0.01&0.16&0.07&10.1\\
\hline
\multicolumn{14}{c}{RR1 template with UVrise as the only AGN template, as in \S5.2.1}\\
$-$0.01&0.18&0.07&13.7& &  0.004&0.21&0.07&21.2& &  $-$0.01&0.16&0.07&10.1\\
\hline
\multicolumn{14}{c}{RR2 template with UVrise as the only AGN template, as in \S5.2.1}\\
$-$0.01&0.17&0.07&12.4& &  0.01&0.19&0.08&17.3&&   $-$0.01&0.16&0.07&10.1\\
\hline
\multicolumn{14}{c}{UV drop behaviour in place of UVrise, as in \S5.2.1}\\
0.06&0.52&0.07&14.3& &  0.21&0.88&0.08&23.1& &  $-$0.01&0.16&0.07&10.1\\
\hline
%\multicolumn{14}{c}{With the Red AGN as well as RR1 and RR2, as in \S5.2.1}\\
%$-$0.01&0.17&0.07&12.4& &  0.002&0.19&0.08&17.3& & $-$0.01&0.16&0.07&10.1\\
%\hline
\multicolumn{14}{c}{The 4 CWW templates replacing E, Sab, Sbc, Scd, and Sdm templates, as in \S6.2}\\
0.01&0.36&0.07&14.3&  & $-$0.05&0.20&0.08&21.1& & 0.04&0.42&0.06&11.0\\
\hline
\multicolumn{14}{c}{Sab, R1UVrise and RR2UVrise as the only templates, as in \S6.2}\\
0.08&0.57&0.07&36.6&  &  0.39&0.96&0.06&59.6& &  $-$0.07&0.21&0.08&25.7\\
\hline
\multicolumn{14}{c}{No AGN templates, as in \S5.2.1}\\
$-$0.07&0.31&0.07&21.1& &  $-$0.18&0.49&0.08&44.2& &  $-$0.01&0.16&0.07&10.1\\
\hline
\multicolumn{14}{c}{A$_{v}$ free for AGN templates, as in \S5.2.2}\\
0.08&0.50&0.06&23.0& &  0.25&0.86&0.05&50.0& &  $-$0.01&0.16&0.07&10.1\\
\hline
\multicolumn{14}{c}{A$_{v}$=0 to 3, with $\beta=3$ in A$_{v}$ prior (see \S\ref{subsec:code})}\\
$-$0.02&0.18&0.08&13.7&& 0.01&0.20&0.08&17.3&& $-$0.03&0.16&0.07&11.9\\
\hline
\multicolumn{14}{c}{A$_{v}=-0.4$ to 1.0}\\
0.03&0.37&0.07&13.7&& $-$0.01&0.20&0.07&17.3& & 0.05&0.43&0.07&11.9\\
\hline
\multicolumn{14}{c}{A$_{v}$ set to zero for all templates, as in \S5.2.2}\\
0.08&0.43&0.08&17.4& &  0.22&0.72&0.09&26.9& &  0.01&0.16&0.08&12.8\\
\hline
\multicolumn{14}{c}{AGN prior removed, as in \S5.2.2}\\
$-$0.03&0.19&0.07&13.7& &  $-$0.08&0.23&0.09&19.2& &  $-$0.01&0.16&0.07&11.0\\
\hline
\multicolumn{14}{c}{AGN prior set to 2 instead of 4, as in \S5.1.3}\\
$-$0.02&0.19&0.07&14.3&& $-$0.03&0.23&0.08&21.1& &$-$0.01&0.16&0.07&11.0\\
\hline
\multicolumn{14}{c}{AGN prior set to 25 instead of 4, as in \S5.1.3}\\
0.06&0.45&0.07&16.8&& 0.21&0.76&0.08&28.8&& $-$0.01&0.16&0.07&11.0\\
\hline
\multicolumn{14}{c}{Brighter AGN limits of --29$<$M$_B$$<$--17.5, as in \S5.2.2}\\
0.13&0.68&0.07&14.9& &  0.43&1.17&0.07&25.0& &  $-$0.01&0.16&0.07&10.1\\
\hline
\multicolumn{14}{c}{Reducing faint AGN limits to --27.0$<$M$_B$$<$--15, as in \S5.2.2}\\
$-$0.01&0.19&0.07&13.7& &  $-$0.02&0.24&0.08&21.2& &  $-$0.01&0.16&0.07&10.1\\
\hline
\multicolumn{14}{c}{Brighter galaxy limits to --23.5$<$M$_B$$<$--13.5, as in \S5.2.2}\\
0.04&0.37&0.07&13.0& &  0.02&0.17&0.07&15.4& &  0.04&0.44&0.07&15.4\\
\hline
\multicolumn{14}{c}{Brighter galaxy limits to --25.5$<$M$_B$$<$--13.5, as in \S5.2.2}\\
0.12&0.53&0.07&18.6& &  0.05&0.20&0.07&19.2& &  0.16&0.63&0.07&18.3\\
\hline
\multicolumn{14}{c}{Reducing faint galaxy limits to --23$<$M$_B$$<$--17.5, as in \S5.2.2}\\
0.04&0.35&0.07&12.4& &  0.02&0.19&0.08&15.4& &  0.15&0.41&0.07&11.0\\
\hline
\hline
\end{tabular}
\footnotesize
\label{table:cdfnresults}
\normalsize
\end{table*}
%%%%%%%%%%%%%%%%%%%%%%%%%%%%%%%%%%
\subsection{ImpZ summary parameters}
\label{subsec:photzsummary}
The final best-case setup for application of ImpZ to the INT WAS ELAIS as a whole is as follows:

Templates -- 6 galaxy templates (E1, Sab, Sbc, Scd, Sdm and Starburst) and 2 AGN templates (RR1UVrise and RR2UVrise), with IGM and Galactic extinction corrections.

Fitting -- AGN templates fit to $stellar$ sources only and galaxy templates fit to all sources, with the prior that a successful galaxy fit must have $\chi^{2}_{red}.^{galaxy}<(\chi^{2}_{red}.^{AGN}-4)$ if the source is $stellar$.  A$_v$ freedom is allowed for Sab, Sbc, Scd, Sdm and Starburst templates, with a prior that makes low A$_v$ fits preferable.

Limits -- A$_v$ limits of 0.0 to 1.0 and absolute magnitude limits of [--22.5--$2log_{10}(1+z)]<$M$_B$$<$--13.5 for galaxies and --27.0$<$M$_B$$<$--17.5 for AGN.
%%%%%%%%%%%%%%%%%%%%%%%%%%%%%%%%%%%%%%%%%%%%%
\section{Error analysis}
\label{sec:error}
Errors in the derived photometric redshifts arise from several causes.  The first is the inherent error in the measured flux of sources, which is expected to be increasingly important for fainter sources.  The template-fitting technique adds further error due to fitting the continuum of observed galaxy SEDs with a set of standard templates which can only sample this continuum.  This is known as cosmic variance.

Catastrophic errors are usually due to degeneracies in the colour--redshift--extinction space.  For example, a late-type, heavily extincted galaxy may appear similar to an early-type unreddened galaxy, but have a very different redshift.  The relative magnitude of these sources of error needs to be more properly quantified.
\subsection{Photometry}
\label{subsec:photerror}
%See Massaroti ..01 B For good ERROR STUFF\\
In order to remove the effect of cosmic variance from the analysis, a synthetic catalogue of fluxes ($U$, $g\arcmin$, $r\arcmin$, $i\arcmin$ and $Z$) was generated 5 times, for 7,000 hypothetical $sources$.  These $sources$ were created by redshifting the templates (equal numbers of each type) and calculating their fluxes in each band.  The redshift distribution was chosen to resemble that of `main' galaxies in the SDSS First Data Release, with a simple Gaussian distribution peaked at $-$20 with a $\sigma$ of 7 to model the absolute magnitude (r$\arcmin$) distribution of the $sources$ (additionally, $sources$ were prevented from having extreme absolute magnitudes).  The fluxes were randomly altered in a Gaussian distribution of errors of width determined by the 1$\sigma$ values of the photometric errors in the INT WAS ELAIS catalogue plus 1 per cent of the flux.  This Monte Carlo simulation generates 35,000 different $sources$.  The photometric redshift code is then used to predict, for A$_{v}$=0, the redshifts and spectral types of the $sources$.  Since the synthetic catalogue is constructed from the same set of templates as those used in the ImpZ code, cosmic variance is eliminated, leaving only the effect of photometric error.

\begin{figure*}
\begin{center}
\includegraphics[height=14.5cm,width=8.5cm,angle=90]{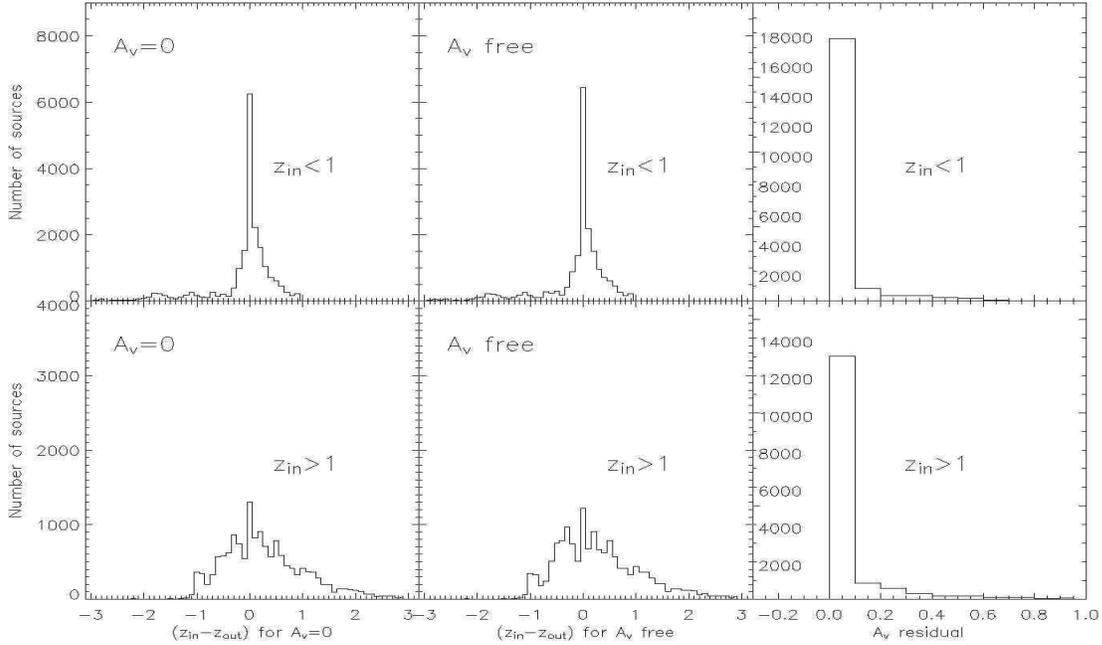}
\caption{\scriptsize{Results of Monte Carlo synthetic catalogue investigations with 35,000 $sources$ (Section \ref{sec:error}.  Top row shows results for `low' input redshift $sources$ (those with z$_{in}<$1), set out as follows: first plot is the distribution of residuals between z$_{in}$ and z$_{out}$), for template-fitting by ImpZ with A$_{v}$ freedom turned off; second plot is the same, but for solutions with A$_{v}$ freedom allowed in ImpZ fitting, as described in \ref{subsec:photzsummary}; third plot shows the A$_{v}$ residual from fitting with A$_{v}$ freedom turned on.  This is simply the A$_{v}$ of the fit since the input synthetic catalogue was generated without A$_{v}$.  The second rows displays the same information for the `high' input redshift sources (those with z$_{in}>$1).}}\label{fig:phot_err}
\end{center}
\end{figure*}

Fig. \ref{fig:phot_err} shows the result of this procedure.  The first plot (both rows) shows the distribution of residuals between z$_{in}$ and z$_{out}$, with the first row showing results for low-redshift input sources ($z_{in}<1$) and the second row showing results for high-redshift input sources ($z_{in}>1$).  The distribution of error in redshift for the low-z$_{in}$ sample is very strongly peaked at zero and is relatively symmetric about this, but the low lying extension to larger errors is not that of a purely Gaussian distribution.  These large shifts in the redshift occur when a secondary minimum in the $\chi^{2}_{red}$ distribution for the $source$ fluxes becomes the primary minimum when the $source$ fluxes are perturbed.  %There is a clearly a degeneracy in the redshift--template space causing a (z$_{in}$-z$_{out}$)$\sim-$1 bump.  For $sources$ at z$_{in}<$1, neither the Lyman limit or Ly$\alpha$ template features have yet entered any of the filters.  Hence a large part of the fit rests on the correct identification of the Balmer break, O[III] doublet, H$\alpha$ or the overall template shape itself.  If it is a confusion between different spectral features then the two spectral features need to be separated by around 1 in redshift space when they enter the same filter in order to generated a residual of $\sim$1.  The degeneracy could therefore be confusion between the Balmer break and H$\alpha$ in the Z band band, or possibly even in the i$\arcmin$ (Table \ref{table:features}).  
Hence for the low-z$_{in}$ sample it can be seen that for the large majority of $sources$, the redshift solution is close to the correct value, with only a low-level Gaussian spread due to photometric error.

For the high-z$_{in}$ sample the distribution is also quite symmetrical about zero, but is less strongly peaked.  This is to be expected due to the poorer signal-to-noise photometry of these more distant sources.  For the high-z$_{in}$ sample, then, the dominant effect is the error in the photometry, causing the general broadening in the residual distribution.

The spectral types of the simulated galaxies are compared with the predicted types in Table \ref{table:types}.  This gives the percentage of galaxies for which their spectral type in the synthetic catalogue was successfully reproduced for each template.\\
From this analysis we conclude that without A$_v$ freedom, the simulated $sources'$ redshifts are well reproduced within 0.1 in (z$_{in}-$z$_{out}$) for $0<z<1$ and roughly 0.4 in (z$_{in}-$z$_{out}$) for $z>1$.  Also, for low redshift $sources$, photometric errors have a reduced effect on the results, whereas for high redshift $sources$ the precision of the photometry has the major influence on the redshift accuracy.
%%%%%%%%%%%%%%%%%%%%%%%%%%%%%%%%%%%%%%%%%%%%%%%%
%%%%%%%%%%%%%
%%%%%%%%%%%%%%%%%%%%%%%
\subsection{Templates}
\label{subsec:templateerror}
In order to investigate the effect of the choice of template-set on the photometric redshifts the 4 UV-extended CWW templates (E, Sbc, Scd, Im) are used in place of 5 galaxy templates in the ImpZ code (E1, Sab, Sbc, Scd, Sdm), retaining the starburst template and the two AGN templates, RR1UVrise and RR2UVrise.  The code is then applied to INT WAS ELAIS and CDFN spectroscopic catalogues.  The resulting photometric redshifts are compared to the output when the original templates were used.  The results are almost as good with $\sigma_{tot}$=0.13, $\sigma_{red}$=0.07 and $\eta$=5.6 per cent for the INT WAS sample and $\sigma_{tot}$=0.36, $\sigma_{red}$=0.07 and $\eta$=14.3 per cent for the CDFN.  The 6 galaxy templates were also replaced with just the Sab template, in order to represent an undersampling of SED space.  It was found that even in this rather extreme example, $\sigma_{tot}$=0.14, $\sigma_{red}$=0.07 and $\eta$=8.6 per cent for the INT WAS sample and $\sigma_{tot}$=0.57, $\sigma_{red}$=0.07 and $\eta$=36.6 per cent for the CDFN.

It is clear that the template-fitting procedure is largely unaffected by the specific choice of templates to use, provided that the main SED features of normal galaxies (such as the Balmer break) are represented.

Table \ref{table:types} shows the comparison between the input $source's$ template and the template that was fit for it for the 35,000 $sources$ generated in \S\ref{subsec:photerror}.  It is immediately clear that the majority (75 per cent) of the input template types are recovered, and that there is almost no degeneracy between early and late--type templates.  There is also a clear demarcation between the galaxy templates and the AGN templates -- very few AGN templates are fit as galaxy templates (with some overlap with the starburst template).  Recall that the reverse does not occur (galaxy template fit by AGN) since $sources$ with an input AGN template are treated as $stellar$ by ImpZ.  Although there are some degeneracies between the templates, this does not effect the photometric redshift accuracy greatly (as can be seen by the redshift residuals in Fig. \ref{fig:phot_err}).  Hence the photometric redshift of a galaxy is more certain than its exact spectral type, as expected if the bulk of redshift identification is due to common features such as the Balmer break and in agreement with the results of fitting the INT WAS ELAIS catalogue with only one galaxy template, and also from the study of \cite{Bolzonella2000A&A...363..476B}.
 %%%%%%%%%%%%%%%%%%%%%%%%%%%%%%%%%%%%%%%
 \begin{table}
 \caption{\scriptsize{Comparison between the input model template types and the best-fitting templates when ImpZ was run on the Monte--Carol generated model catalogues.  Values are given as the percentage for each input template that was fit by each output template.  Ideally only the diagonal region (bold) would be populated.}}
\begin{tabular}{|c c c c c c c c c|}
\hline
\hline
Out&\multicolumn{8}{c}{Input SED type} \\
\hline
& \scriptsize{E} & \scriptsize{Sab}& \scriptsize{Sbc}& \scriptsize{Scd} & \scriptsize{Sdm} & \scriptsize{Sbrst} & \scriptsize{RR1} & \scriptsize{RR2}\\
  \hline
\scriptsize{E} &    \bf{88}      &  0.0  &       0.0    &     0.0    &       0.0   &    0.0  &   0.0  &   0.0\\
\hline
\scriptsize{Sab} &  8.0   &    \bf{71.7}   &     18.6    &     0.0    &      2.9   &     0.0  &  0.1     &  0.3\\
\hline
\scriptsize{Sbc} & 2.0    &    8.3    &    \bf{71.2}        &  0.0 &         0.0    &  0.0  &    0.0      & 0.0 \\
\hline
\scriptsize{Scd} &2.0   &  8.3      &     10.2     & \bf{84.8}    &      4.3     &   7.4   &   0.0    &  0.0\\
\hline
\scriptsize{Sdm} &  0.0    &  10.0     &   0.0    &  3.0     & \bf{87}      & 25.0  & 0.0     & 0.0\\
\hline
\scriptsize{Sbrst} &     0.0    &  1.7      &  0.0     & 12.2    &  5.8   &    \bf{67.6}    &  3.9      &  4.7\\
\hline
\scriptsize{RR1} &  0.0  &  0.0   &    0.0 &  0.0      &       0.0       &      0.0     &      \bf{64.3}  &    30.3\\
\hline
\scriptsize{RR2} &  0.0           &      0.0     &       0.0  & 0.0  &          0.0         &    0.0   &       31.7     &   \bf{64.7}\\
\hline
\hline
\end{tabular}
\label{table:types}
\end{table}
%%%%%%%%%%%%% 
%%%%%%%%%%%%%%%%%%%%%%%
\subsection{Filters}
\label{subsec:filtererror}
What are the relative importance of different filters?  In order to successfully derive redshifts for different sources, the filter set in use needs to have sufficient wavelength coverage to encompass the main template features for a broad range of redshifts.  As seen in \S\ref{subsubsection:wfsoutliers} and \S\ref{subsubsection:cdfnoutliers}, most outliers were sources whose spectroscopic redshifts meant that the main template features (Balmer break and Lyman limit) did not fall within filter bandpasses.  It is of interest to see how results change when a filter is not included in the template-fitting.  To this end, ImpZ was run on the INT WAS ELAIS catalogue without $U$ band information.  %, and similarly without Z band photometry.  
The statistical results of this can be found in Table \ref{table:wfsresults} but the main effect was that without $U$ band photometry, all z$_{phot}$ values were below 1.5, although 7 sources lie at $z_{spec}>1.5$.  This is because information on the position of the Ly$\alpha$ line has been lost, and at these z$_{spec}$ redshifts no other main template feature falls into the INT WAS bands.  

%Following the same argument, the loss of Z band photometry should not have too large an effect on results, since the other filters can still locate either the Lyman limit, Ly$\alpha$ 
Additionally, a synthetic catalogue of 5000 $sources$ was generated as in \S\ref{subsec:photerror} but without U band photometry.  For the $z_{in}<1$ sample, the $(z_{in}-z_{out})$ residual distribution was strongly peaked at zero, in a similar fashion to the first plot in Fig. \ref{fig:phot_err}, but with a population of outliers with $(z_{in}-z_{out})\approx-2.6$.  This suggests that without the use of the $U$ band to decide if the Lyman break feature is present or not, the Balmer break can be mistaken to be Ly$\alpha$ in the $g\arcmin$ band.  Hence the low-z$_{in}$ $source$ is then placed at a high-z$_{out}$.  As a whole, though, the loss of $U$ band information does not adversely effect the z$_{in}<$1 sample.  In contrast, the $z_{in}>1$ sample is highly dependent on it for a reasonable z$_{out}$ value.  The $(z_{in}-z_{out})$ residual distribution is almost flat -- the z$_{out}$ accuracy is very poor.  Thus the $U$ band is central to good photometric redshift accuacy for $z\ga1.5$.  This is clear from Table \ref{table:features} -- without $U$ band none of the template features listed is present in the remaining INT WAS filters until $z>2.4$ when Ly$\alpha$ enters the $g\arcmin$ band.
%%%%%%%%%%%%%%%%%%%%%%%
\subsubsection{Degeneracies in parameter space}
As seen in \S\ref{subsec:photerror} and \S\ref{subsec:templateerror} there do exist degeneracies in the parameter space and this is a cause of outliers in the photometric error analysis of the simulated catalogues.  Common degeneracies include confusion between the main template features (when either the redshift or the lack of bands means most features don't fall into the filter bandpasses) and confusion between different galaxy templates.  Often, however, the spectral types can be degenerate without altering z$_{phot}$ greatly (compare Table \ref{table:types} with Fig. \S\ref{fig:phot_err}).

It is also of interest to see if A$_{v}$ introduces its own degeneracies.  Such degeneracies will be inherent even if A$_{v}$ is not explicitly fit, since real galaxies will have different A$_{v}$.  In order to quantify this, the same synthetic catalogue procedure as in \S\ref{subsec:photerror} is carried out, but the ImpZ code is allowed to fit for free A$_{v}$.  Since the input catalogue is created with A$_{v}$=0 we can see how fitting for free A$_{v}$ alters the findings.

Fig. \ref{fig:phot_err} shows the result of this procedure as the second and third plots on each row (again the first row is for the $z_{in}<1$ sample and the second row is for the $z_{in}>1$ sample).  The distribution of residuals between z$_{in}$ and z$_{out}$ are very similar to those for the A$_{v}$=0 case,  with both the low-z${in}$ and high-z$_{in}$ samples being strongly peaked at zero.  The low-z$_{in}$ sample has a noticeable but low-level tail to higher residuals, whilst the high-z$_{in}$ sample can be well represented by a broad Gaussian.  The high-z$_{in}$ sample has a slight secondary bump at $(z_{in}-z_{out})\approx-0.4$, which is likely to be due to confusion between the Balmer break and O[III] doublet, since, as can be seen in Table \ref{table:features} , these features enter and leave the i$\arcmin$ and Z band with a separation of around 0.4 in redshift.

The A$_{v}$ residual is also shown.  This is simply the A$_{v}$ of the fit since the input synthetic catalogue was generated without A$_{v}$.  The low-z$_{in}$ sample is very strongly peaked at zero, and the high-z$_{in}$ sample is also strongly peaked at zero, with slightly more A$_{v}>0$ solutions.  This means that the increasing photometric error for more distant/faint $sources$ causes a few $sources$ to be fit as redder $sources$ than they really are.  This however does not necessarily cause an increase in the redshift error, as can be seen from the redshift residual plots for the A$_v$--free fitting.  It is clear that the addition of A$_{v}$ freedom introduces little degeneracy, and the fact that the redshift solutions tend to improve lends weight to the argument that the A$_v$ thus determined reflects the physical A$_v$ of the galaxy.

Running the Monte Carlo simulations with larger photometric errors tended to increase the width of the residual distributions.
\label{subsubsec:degeneracy}
\subsection{Overall errors}
\label{subsec:toterror}
The main sources of error are due to photometry, cosmic variance and degeneracies in colour--redshift--extinction space.  The relative effect of each depends on the signal-to-noise of the source with photometric errors dominating those with low signal-to-noise.  Splitting into low ($z<1.5$) and high redshift ($z>1.5$), we can broadly say that the error for low-redshifts is 0.1 or less in (1+z) whereas for the high-redshifts the error can be as large as 0.3 or more.
%%%%%%%%%%%%%%%%%%%%%%%%%%%%%%%%%%%%%%%%%%%%%%%%%%%%%%%%%%%
\section{Discussion}
\label{sec:disc}
A comparison can be made to the photometric accuracies of other recent works:  photometric redshifts for galaxies in the Great Observatories Origins Deep Survey (GOODS; \citealt{Dickinson2001AAS...198.2501D}) were estimated for 434 galaxies by \cite{Mobasher2004ApJ...600L.167M} where in total there were as many as 18 independent photometric measurements for each galaxy.  Using the Bayesian method (\citealt{Benitez1999prdh.conf...31B}) they obtained a total $rms$ scatter, $\sigma_{tot}$, of 0.11, with an outlier-clipped $rms$, $\sigma_{red}$, of 0.047 and an outlier fraction, $\eta$, of roughly 10 per cent.  This is very similar to the success of the ImpZ code on the INT WAS ELAIS sample, which had a maximum of only 5 photometric bands.   In \cite{Richards2001AJ....122.1151R}, using statistics based on $\Delta_z=(z_{phot}-z_{spec})$, redshifts were estimated for 2625 quasars in the SDSS filter system with an overall $rms$, $\sigma_{z}$ of 0.676 and $\sigma_{red}$ of 0.1 for sources with $|\Delta z|<0.3$.

The success of the ImpZ code can also be  directly compared for quasars to the results of \cite{Barger2002AJ....124.1839B} who got one third of broad-line sources within 25 per cent of z$_{spec}$.  ImpZ fit two thirds of the same CDFN broad-line sample within 25 per cent.  More recently, \cite{Kitsionas0309628} estimated photometric redshifts for a sample of X--ray selected QSOs in the SDSS bands (up to 5 filters), obtaining z$_{phot}$ within $|\Delta z|<0.3$ for 20 of the 30 QSOs at $z>0.4$.  The same statistic can be calculated for INT WAS ELAIS `QSO' $z>0.4$ sub-sample, where 10 of 20 sources are within the same $\Delta$z tolerance, and for the broad-line and/or compact CDFN sources at $z>0.4$, where 36 of 50 (72 per cent) of sources are within $|\Delta z|<0.3$.  The ImpZ quasar redshift results are clearly highly successful and show that reliable redshift estimates for quasars can be achieved.

It is important to note that the intrinsic variability of quasars offsets the magnitudes measured in various bands depending on the actual epoch of observation.  This is certainly an issue for the INT WAS survey which was completed over several years.  The variability scrambles the real spectrophotometric data taken over an extended period whereas the template fits assume that the the photometry in each band is taken at the same time (or that there is no variability) and is therefore sampling the same underlying SED.  This problem was noted in \cite{Wolf2001A&A...365..681W} who found around half their quasar sample had completely incorrect photometric redshifts, presumably mainly due to this problem.  Clearly future surveys need to be scheduled with this effect in mind.

The SWIRE survey (in progress) combines both ground-based optical photometry and IR data (3.6, 4.5, 5.6, 8, 24, 70 and 160$\micron$).  Virtually all SWIRE galaxies are expected to be detected in the IRAC 3.6$\micron$ band.  Since little contribution from dust is expected at this wavelength, this band can be used to improve the photometric redshift estimates.  Additionally, at least for low redshift galaxies, an estimate can be made of the total stellar mass in the galaxy.

Where galaxies are detected at 4.5, 5.8 and 8$\micron$ an estimate can also be made of the luminosity in the cirrus emission component, after subtraction of the predicted starlight contribution.  The ratio of L$_{cirr}$ to, say, L$_B$ should be related to the A$_v$ value.

%If a galaxy is detected at 24$\micron$ and beyond then it will be possible to make estimates of the starburst and AGN dust torus contributions.
%
%
The crucial improvement of a survey such as SWIRE over deep fields such as HDF will be the far larger volume sampled, so that cosmic variance can be properly addressed.  Although photometric redshift techniques have been applied successfully to the HDF, their true statistical power comes from application to large area surveys such as SWIRE, which have large numbers of galaxies.  Application of the ImpZ code to the SWIRE survey will be presented in a future work.
%%%%%%
\section{Conclusions}
\label{sec:conc}
In this work, photometric redshifts have been studied using SED template-fitting on two spectroscopic redshift catalogues (INT WAS ELAIS and CDFN).  The overall accuracy of the redshift code, ImpZ, is found to be good, with $\Delta z/(1+z)<0.1$ for 92 per cent of galaxies (where this statistic is calculated from the 138 `GALAXY' sources in the INT WAS ELAIS catalogue).  The addition of A$_{v}$ freedom in the fitting improves the redshift solution, suggesting that some information about the true A$_v$ of the source is also returned.  Application to sources with known A$_v$ will help to quantify this information (Babbedge 2004; PhD thesis, in prep.).  The template-fitting method is also extended to quasars via the inclusion of AGN templates and is found to be reasonable, with $\Delta z/(1+z)<0.25$ for 68 per cent of quasars (based on the 25 `QSO' sources in the INT WAS ELAIS catalogue).  It is noted that the inherent time-variability of quasars reduces the effectiveness of the technique for surveys where the photometry is collected over an extended period.

A combined galaxy--quasar approach to template-fitting photometric redshift techniques has been presented and the results and analysis presented in this work clearly show that photometric redshifts can be calculated with good success for $both$ galaxies and quasars, as long as certain considerations are taken into account:  it is important to limit the absolute magnitude of the solutions to prevent unphysical results; use available optical morphology/stellarity information in order to pre-select sources that are more likely to be quasars than galaxies (since we expect quasars to be point-like whereas most galaxies will be extended in some way).  Although the application of template-fitting to a single source in order to derive its redshift can never be completely relied on, due to the inherent degeneracies and limitations of a method reliant on perhaps four or five data-points, the statistical information drawn from application to a large catalogue of objects is extremely powerful and can be used for many different investigations.

The ImpZ code will be applied to the entire re-calibrated INT WAS ELAIS N1 and N2 data in order to investigate the evolution of extinction and star formation rates in a companion to this paper; Babbedge et al. (2004; in prep.).
%%%%%%
\section{Acknowledgements}
\label{sec:ack}
We thank an annoymous referee for insightful comments that have served to strengthen this paper.  We wish to thank Jacopo Fritz, who made the major contribution to the development of the spectrophotometric synthesis code adopted to produce the optical--NIR
templates.  We are also grateful Duncan Farrah for his helpful comments.

The INT WAS data is made publically available through the Isaac Newton Groups' Wide Field Camera Survey Programme.  The  Isaac Newton Telescope  is operated on the island of La Palma by the Isaac Newton Group in the Spanish Observatorio del Roque de los Muchachos of the Instituto de Astrofisica de Canarias.  Funding for the creation and distribution of the SDSS Archive has been provided by the Alfred P. Sloan Foundation, the Participating Institutions, the National Aeronautics and Space Administration, the National Science Foundation, the U.S. Department of Energy, the Japanese Monbukagakusho, and the Max Planck Society.  The SDSS Web site is http://www.sdss.org/.  We also thank A. J. Barger and co--workers for making their CDFN catalogue publicly available and M.D. Gregg and M. Lacy for providing the optical and IR spectrum of a red quasar.

MP acknowledges support from the $SPITZER$ Legacy Science Program, funded through support provided by NASA and issued by the Jet Propulsion
Laboratory, California Institute of Technology.  IPF was supported in part by grant PB1998-0409-C02-01 of the Spanish Ministerio de Ciencia y Tecnolog\'\i a.  EAGS acknowledges support by the Marie Curie Fellowship MCFI-2001-01809.

\bibliography{/Users/tsb1/Documents/Work/References/BibdeskBibliog}
\bibliographystyle{mn2e}

\label{lastpage}
\end{document}